\documentstyle[preprint,prb,aps,psfig]{revtex}

\tighten

\begin{document}

\title{What are the experimentally observable effects of vertex corrections 
in superconductors?}

\author{ ${\rm P. Miller}^{\dag}$ \and 
${\rm J. K. Freericks}^{\dag}$ \and ${\rm E. J. Nicol}^{\ddag}$ }

\address{
$^{\dag}$ Department of Physics, Georgetown University, \\
 Washington, D.C. 20057-0995, U.S.A.\\
$^{\ddag}$ Department of Physics, University of Guelph, Guelph, Ontario, Canada }

\maketitle
\begin{abstract}
{\sloppy
We calculate the effects of vertex corrections, of non-constant density of
states and of a (self-consistently determined) phonon self-energy 
for the Holstein model on a 3D cubic lattice. 
We replace vertex corrections with a Coulomb pseudopotential, $\mu^{*}_{C}$, 
adjusted to give the same $T_{c}$, and repeat the calculations, to see which 
effects are a distinct feature of vertex corrections. 
This allows us to determine directly observable effects of
vertex corrections on a variety of thermodynamic properties of superconductors. 
To this end, we employ conserving 
approximations (in the local approximation) to calculate the superconducting 
critical temperatures, isotope coefficients, superconducting gaps, free-energy 
differences and thermodynamic critical fields for a range of parameters. 
We find that the dressed value of 
$\lambda$ is significantly larger than the bare value. 
While vertex 
corrections can cause significant changes in all the above quantities (even when
the bare electron-phonon coupling is small), the changes can usually be well-modeled 
by an appropriate Coulomb pseudopotential. 
The isotope coefficient proves to be the quantity that 
most clearly shows effects of vertex corrections that can not be 
mimicked by a $\mu^{*}_{C}$. } 

\end{abstract}

\vskip1.0in


\vfill

\section{Introduction}

The theory of conventional, low-temperature superconductors has been well 
understood for decades, within BCS theory~\cite{BCS}, 
and many material properties of 
superconductors have 
been accurately described within the more appropriate Migdal-Eliashberg 
formalism~\cite{Migd,Elia1}, 
which includes the retardation effects of the 
electron-phonon interaction in a realistic manner. The success 
of the formalism arises from Migdal's theorem~\cite{Migd}, 
which says that vertex corrections can be neglected when the ratio of the 
phonon energy scale to the electron energy scale is small 
(such as the value of $10^{-4}$ 
typical of conventional low-temperature superconductors). 
The 
physical reason being that the ion movement is typically too slow to respond to 
anything but the mean-field potential produced by the fast-moving electrons. 

In recent years there has been investigation of superconducting 
materials~\cite{Matt1,Cava,Vash,Matt2,Heba,Holc,Ross} 
such as Ba$_{1-x}$K$_x$BiO$_3$, 
K$_3$C$_{60}$ and the A$15$s, 
whose phonon energy scale is a larger fraction of 
their electron energy scale. 
For such materials, the standard Migdal-Eliashberg theory may no longer be valid, 
either because second-order processes in the electron-phonon coupling 
(so-called vertex corrections) 
can not be 
neglected~\cite{Free1,Free2,Free3,Nico,Free4}, or because 
structure in the electronic density of states and finite-bandwidth effects become 
significant~\cite{Mars}. Our aim is to identify those experiments which can 
clearly indicate where vertex corrections, or structure in the electronic density 
of states, are observable in real materials. In particular, we wish to 
uncover those effects which can not be mimicked by a Coulomb pseudopotential, 
$\mu^{*}_{C}$, as 
this would render them unobservable in practice, because a $\mu^{*}_{C}$ is 
typically fitted to the experimental data.

The most compelling evidence for the effect of vertex corrections would come 
from a tunneling conductance measurement that provided data for the
tunneling conductance out to an energy at least twice that of the maximal
phonon frequency of the bulk material.  Then a tunneling inversion could be
performed using only the experimental data out to an energy of the maximal 
bulk phonon energy, and the results for Migdal-Eliashberg theory
would be compared to the vertex-corrected
theory for the experimental data that was measured in the 
multiphonon region, at voltages above the maximal bulk phonon energy
of the material.  Unfortunately, such an analysis has only been performed for
lead\cite{Free2} and in that case, the experimental data was not accurate enough
in the multiphonon region to be able to see if effects of vertex corrections
were observable. Similarly, high energy data from optical conductivity experiments 
would indicate whether or not vertex corrections are important, but there again, 
the accuracy of the data might preclude seeing effects of vertex corrections.

In order to determine what effects are unmistakenly due to vertex corrections,
we fit a Coulomb pseudopotential $\mu^*_C$ to a Migdal-Eliashberg theory so
that the superconducting transition temperature $T_c$ is the same for the
vertex-corrected theory (with no $\mu^*_C$) and the conventional Migdal-Eliashberg
theory (with a $\mu^*_C$ but no vertex corrections). 
In the case of dressed phonons, we adjust both the electron-phonon 
interaction energy and $\mu^*_C$, in order to fit the same values of $\lambda$, 
and $T_{c}$ as the results with vertex corrections. 
This procedure is exactly
the procedure carried out in analyzing experimental data within the conventional
theory.  We carry out this ``experimental analysis'' on a model 
system where we know that vertex corrections have a large effect, in order to 
see the extent to which the conventional analysis masks their observation. 
Hence we look for differences in thermodynamic quantities between these 
results and those with vertex corrections. Any quantities which are significantly 
different when calculated with a fitted $\mu^*_C$ instead of vertex corrections 
pin-point the experiments that can give the best indication that a material
needs to have vertex corrections included in its description.

\begin{figure}
  \centerline{\psfig{figure=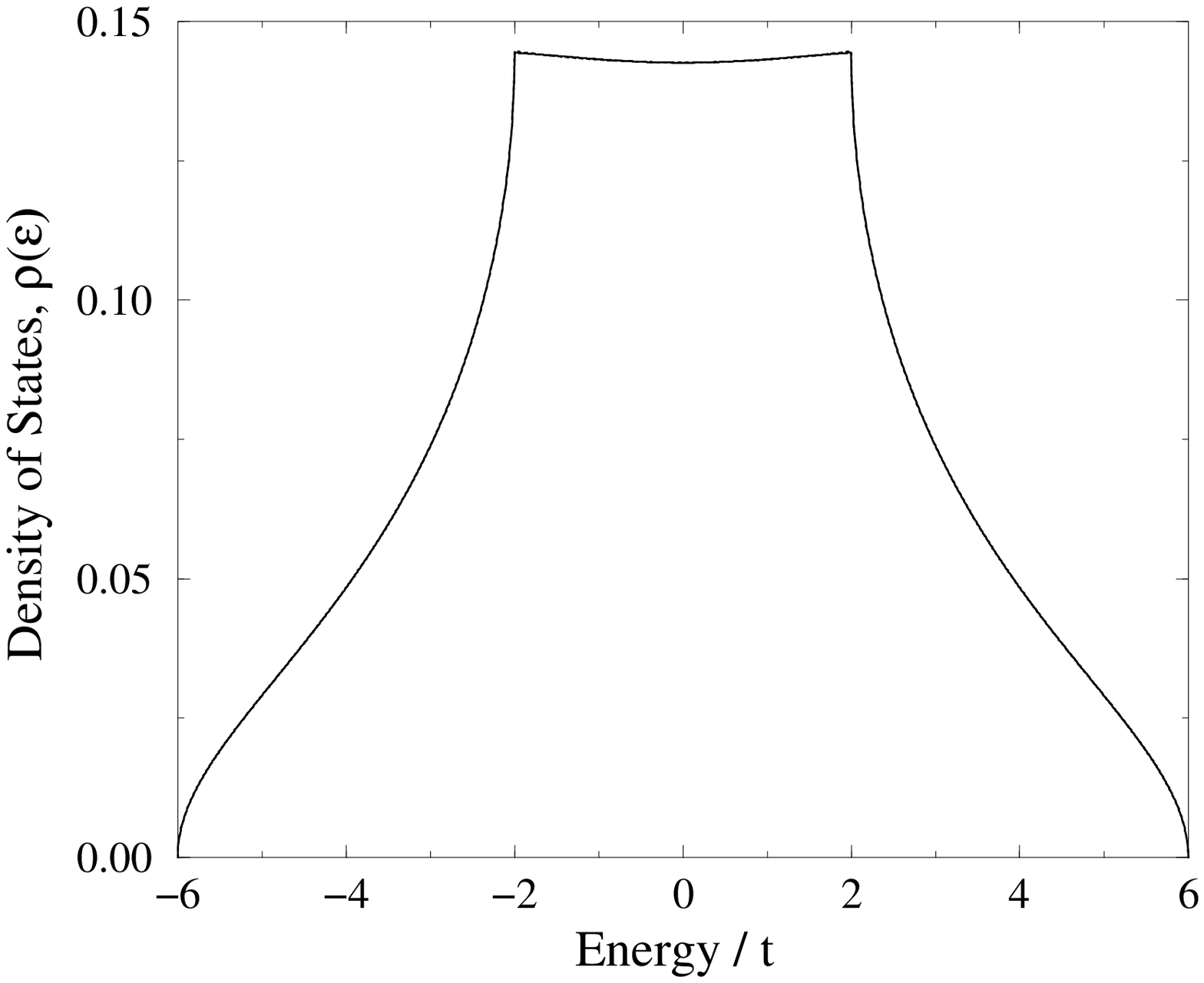,width=4.5in}}
  \caption{The single-spin density of states for non-interacting electrons on a 
   3D cubic tight-binding lattice. 
   Note the nature of the van Hove singularities, which lead to an abrupt 
   fall in the density of states. They occur at electron fillings of 
   $n\approx 0.45$ and $n\approx 1.55$. The approximate form used in the 
   calculations is also plotted, and is indistinguishable by eye from 
   the exact curve.}
\label{fig:dos3d}
\end{figure}

Since this is the initial attempt at solving such a
problem, we study a simple model system, 
which has a non-constant electronic density of 
states (Figure~\ref{fig:dos3d}), using both Migdal-Eliashberg formalism, and going 
beyond it, to include vertex corrections. The vertex corrections are second-order 
diagrams where a pair of phonon lines cross, as shown in the self-energies of 
Figure~\ref{fig:self}(b). In both cases, the phonon propagators can be dressed 
({\it i.e.} with a phonon self-energy included) 
or bare, as shown in Figure~\ref{fig:self}. We find that 
dressing the phonons leads to a strong enhancement of the effective interaction 
strength, exemplified by a large renormalization in the value of the 
electron-phonon coupling parameter, $\lambda$ . 

The specific model we study is the Holstein Hamiltonian~\cite{Hols} 
on a 
3D cubic lattice. The Hamiltonian consists of conduction electrons that hop 
from site to site, coupled to harmonic, localized (Einstein) phonons: 
\begin{equation}
\hat{H} - \mu\hat{N} = 
\sum_{i,j,\sigma} t_{ij} \hat{c}^{\dagger}_{i\sigma} 
\hat{c}_{j\sigma} + 
\frac{1}{2}M\Omega^{2} \sum_{i} \hat{x}_{i}^{2} + 
\frac{1}{2M}\sum_{i} \hat{p}_{i}^{2} +
\sum_{i} (g\hat{x}_{i} - \mu)\hat{n}_{i} ,
\label{eq:hhol}
\end{equation}
where $\hat{c}^{\dagger}_{i\sigma}$ and $\hat{c}_{i\sigma}$ are fermionic 
operators which create and destroy, respectively, an 
electron of spin-$\sigma$ in a single Wannier (tight-binding) state on the
 lattice site {\it i}, whose total electron occupancy is given 
by $n_{i} = \hat{c}^{\dagger}_{i\uparrow} \hat{c}_{i\uparrow} + 
 \hat{c}^{\dagger}_{i\downarrow} \hat{c}_{i\downarrow}$. 
The electron hopping is between nearest neighbors only, such that 
$t_{ij} = -t $ if $i,j$ are neighboring sites, with $t$ the overlap integral, 
and $t_{ij} = 0$ otherwise. 
The phonons, of mass, $M$, 
with displacement $x_{i}$ and momentum, $p_{i}$, are 
characterized by their frequency, $\Omega$. The strength of 
the bare electron-phonon coupling 
can be measured by the bipolaron binding energy, $U$, where:
\begin{equation}
U = - \frac{ g^{2} }{ M \Omega^{2} } .
\end{equation} 
The chemical potential is $\mu$, 
and is always calculated self-consistently for a given 
average number of electrons per site, $n$ ($0 \leq n \leq 2$). 
Particle-hole symmetry occurs at half-filling, where $n=1$ and $\mu=U$. 
We concentrate our work on superconductivity and ignore any possible 
charge-density wave order that may occur near half-filling.

We carry out weak-coupling expansions within the conserving approximations of Baym and 
Kadanoff~\cite{Baym1,Baym2,DoMa}. 
The electron self-energy, $\Sigma$, is given as a 
functional derivative of the free-energy functional, $\Phi$, with respect to the 
electron Green's function, $G$. When dressing the phonons, we 
maintain a conserving approximation by careful choice of the phonon self-energy. 
In this case, 
there must be a free-energy functional whose {\it partial} derivative 
with respect 
to the electron Green's function yields the electron self-energy, {\it and} whose   
{\it partial} derivative with respect 
to the phonon Green's function, $D$, yields the phonon self-energy. 
In such cases there always exists a related free-energy functional whose {\it full} 
functional derivative with respect to the 
electron Green's function yields the electron 
self-energy. In all cases, the irreducible vertex function is given by the 
{\it full} functional 
derivative of the electron self-energy with respect to the electron Green's function, 
{\it i.e.} $\Gamma = \delta \Sigma  / \delta G$.

We employ the local approximation in our calculations, which means we neglect the 
momentum dependence in the self-energy and irreducible vertex functions. 
 The local problem gives an exact solution in the 
infinite-dimensional 
limit~\cite{Free1,Free6}, but in this case it is an approximation, 
which, as used 
in Migdal-Eliashberg theory, is expected to give good quantitative results, 
though of course it means we can only study s-wave pairing.

We use the formalism, described in detail in the next section, to calculate the 
critical temperature, $T_{c}$ and the isotope coefficient, $\alpha$, by looking at the 
instability of the normal state. Also, within the superconducting state, we calculate 
the superconducting gap, $\Delta$, and the thermodynamic critical field, $H_{c}$, from 
the free-energy difference, $F_{S}-F_{N}$. We point out in particular those results 
which deviate from Migdal-Eliashberg theory, and arise from vertex corrections or a 
non-constant density of states. By including a Coulomb pseudopotential, which 
causes the same change in $T_{c}$ as vertex corrections, we are able to demonstrate 
what experiments can be used to differentiate between vertex corrections and a 
Coulomb pseudopotential. 

Section II contains the formalism, describing the 
particular approximations we use, and explaining 
how our calculations are carried out. 
Section III presents our computational results, 
including a number of graphs depicting the various quantities that can be 
determined experimentally for real systems. 
We make our conclusions in Section IV, which is followed by an Appendix which 
gives the detailed formula withheld from Section II.

\section{Formalism}

\begin{figure}
  \centerline{\psfig{figure=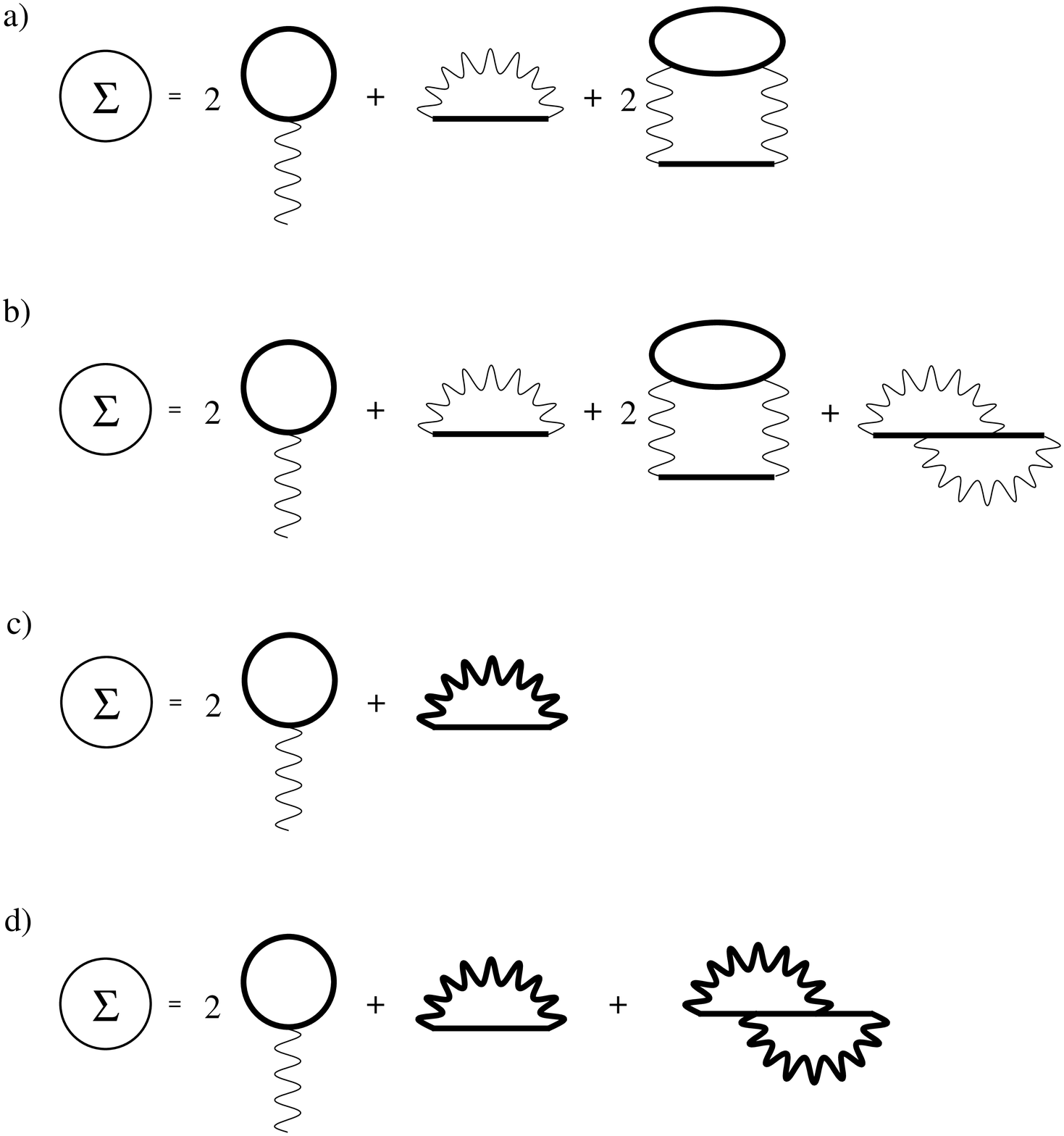,width=6.0in}}
\caption{Different forms for the electron self-energy. All diagrams include the same 
initial Hartree term, followed by the specific Fock contribution and any other 
required diagrams. All electronic Green's functions are dressed, but phonon lines can 
be undressed (thin) or dressed (thick) as in Figure~\ref{fig:dself},b). 
For bare phonons, (a) is with no vertex corrections, while (b) includes the 
vertex-corrected diagram. When the phonons are dressed, (c) is without vertex 
corrections, while (d) includes them. A factor of 2 appears in front of each 
electron loop term, to indicate a sum over spins.}
\label{fig:self}
\end{figure}

In our calculations, we use four different types of conserving approximations, which 
will be expounded below. Four different approximations are necessary, in order to 
reveal how dressing the `bare' phonons affects solutions of the model, as well as 
to demonstrate the effects of vertex corrections. 
Each approximation  includes a specific 
self-energy, and hence a specific vertex function, 
as shown in Figures~\ref{fig:self}-\ref{fig:vert}. 
In short, these can be described as (a)~Migdal-Eliashberg approximation 
with a truncated dressing 
of phonons; (b)~second-order perturbation theory, which contains vertex corrections 
and a truncated dressing of phonons; (c)~Migdal-Eliashberg theory with dressed 
phonons; and (d)~vertex-corrected theory with dressed phonons. 

The calculations are carried out on the imaginary axis~\cite{Owen}, with the Green's 
functions and self-energies defined at Matsubara frequencies, 
$i\omega_{n} = (2n+1)\pi i T$, where $T$ is the temperature. 
The self-energies are calculated from the derivative of an appropriate 
free-energy functional, $\Phi$, as: 
$\Sigma(i\omega_{n}) = \frac{1}{T} \frac{ \delta\Phi}
{\delta G^{\uparrow\uparrow}(i\omega_{n})}$, 
$\phi(i\omega_{n}) = \frac{1}{T} 
\frac{ \delta\Phi}{\delta F^{*}(i\omega_{n})}$, 
while the irreducible vertex function for superconductivity is given by: 
$\Gamma_{n,m} = \frac{ \delta\phi(i\omega_{n})}{\delta F(i\omega_{m})}$. 
Figure~\ref{fig:free} shows the Feynman diagrams which correspond to 
the appropriate free-energy functional, 
$\Phi$, for each of the four calculations, 
while we defer the specific formulae to the 
appendix. 
In the diagrams, taking a functional derivative corresponds to cutting 
and removing the appropriate Green's function line. 

When calculating properties for the superconducting state, we must use the Nambu 
formalism~\cite{Namb}, where the electron Green's function and self-energy 
are represented by a $2 \times 2$ matrix:
\begin{equation}
\underline{\underline{G}}(i\omega_{n}) \equiv 
\left( 
\begin{array}{cc}
G^{\uparrow\uparrow}(i\omega_{n}) & F(i\omega_{n}) \\
F^{*}(i\omega_{n}) & -G^{\downarrow\downarrow}(-i\omega_{n})
\end{array} 
\right)
= \int \rho^{(0)}(\epsilon) d\epsilon 
\left[ i\omega_{n}\underline{\underline{\tau}}_{3} \, - \epsilon + \mu - 
\underline{\underline{\Sigma}}(i\omega_{n}) \right]^{-1},
\label{eq:greeng}
\end{equation}
where $\rho^{(0)}(\epsilon)$ is the approximate form for the 
non-interacting electron density of 
states on a 3D cubic lattice as shown in Figure~\ref{fig:dos3d}, given by 
Uhrig~\cite{Gotz}, and 
\begin{equation} 
\underline{\underline{\Sigma}}(i\omega_{n}) \equiv
\left( \begin{array}{cc}
\Sigma(i\omega_{n}) & \phi (i\omega_{n}) \\
\phi^{*} (i\omega_{n}) & \Sigma^{*} (i\omega_{n})
\end{array} \right), \,
\underline{\underline{\tau}}_{3} = 
\left(\begin{array}{cc} 1 & 0 \\*  
0 & -1 \end{array} \right). 
\end{equation}
Here, the diagonal and off-diagonal Green's functions are defined respectively as:
\begin{eqnarray}
G^{\sigma\sigma}(i\omega_{n}) & = & -\frac{1}{N} \sum_{j} \int^{\beta}_{0} d\tau 
\exp (i\omega_{n}\tau) 
\left<{\rm T}_{\tau} 
\hat{c}_{j\sigma}(\tau) \hat{c}^{\dagger}_{j\sigma}(0) \right>,
\\* 
F(i\omega_{n}) & = & -\frac{1}{N} \sum_{j} \int^{\beta}_{0} d\tau 
\exp (i\omega_{n}\tau) 
\left<{\rm T}_{\tau} 
\hat{c}_{j\uparrow}(\tau) \hat{c}_{j\downarrow}(0) \right>,
\end{eqnarray}
where ${\rm T}_{\tau}$ denotes time-ordering in $\tau$. 
The definitions of $\Sigma(i\omega_{n})$ and $\phi(i\omega_{n})$ for 
each calculation are given in the appendix, 
Eqs.~(\ref{eq:selfa})-(\ref{eq:dselfd}), 
and are represented in Figure~\ref{fig:self}. 

The phonon propagator, $D(i\omega_{\nu})$, is defined by~\cite{Fett}:
\begin{equation}
D(i\omega_{\nu}) = \frac{ m\Omega^{2}}{N} \sum_{j} \int^{\beta}_{0}
d\tau \exp [i\omega_{\nu}\tau]
\left< T \hat{x}_{j}(\tau) \hat{x}_{j}(0) \right>,
\end{equation}
such that the `bare' propagator, with no self-energy, $D^{(0)}(i\omega_{\nu})$, 
is equal to $\frac{- \Omega^{2}}{\omega_{\nu}^{2} + \Omega^{2}}$. The 
appropriate Matsubara frequencies, $\omega_{\nu}$, in this case, are those which 
lead to bosonic statistics, such that $i\omega_{\nu} = 2\pi iT \nu$. 

When we use bare phonons, and have no vertex corrections, the electron self-energy 
is illustrated in Figure~\ref{fig:self}(a). The first term is the Hartree 
contribution, $Un$. It can be included 
by a shift of the chemical potential as it only contributes a constant 
to the diagonal part 
of the self-energy, $\Sigma(i\omega_{n})$. 
Such a shift is implicitly included when $\mu$ 
appears in a Green's function, so that the Hartree term is hereafter neglected. 
It is followed by the Fock term, then a single-phonon dressing term, where the 
phonon line includes a single loop, which is the 
electron polarizability, $\pi^{(0)}(i\omega_{\nu})$:
\begin{equation}
\pi^{(0)}(i\omega_{\nu}) = -2T\sum_{m} \left[ G(i\omega_{m}) G(i\omega_{m+\nu})
-  F(i\omega_{m}) F^{*}(i\omega_{m+\nu}) \right],
\label{eq:chi}
\end{equation} 
with the factor of two arising from the summation over spin.
The third term includes dressing of phonons in a truncated manner, 
and allows us to make comparison with the complete second-order approximation, in 
Figure~\ref{fig:self}(b), 
where the only difference is the inclusion of the vertex-correction 
term. The specific formula for the extra term in Figure~\ref{fig:self}(b), 
which enters our calculations when we 
include vertex corrections using bare phonons, is given in 
Eqs.~(\ref{eq:selfb}) and~(\ref{eq:phib}). 
The extra contributions to the self-energy coming from vertex corrections 
have opposite sign to the Fock term, near  
half-filling, where the product of two electronic Green's functions is negative (the 
Green's functions are pure imaginary at half-filling). 
Away from half-filling, the Green's functions gain real parts, 
which means that near the band edges the product of two Green's functions can be 
positive, and the vertex-correction terms then add to the Fock term.

In the calculations with dressed phonons, we have a Dyson's equation 
for the phonon propagator, as depicted in Figure~\ref{fig:dself}:
\begin{equation}
D(i\omega_{\nu}) =  D^{(0)}(i\omega_{\nu}) + 
D^{(0)}(i\omega_{\nu}) \Pi(i\omega_{\nu}) D(i\omega_{\nu}) ,
\label{eq:greend}
\end{equation}
where $ \Pi(i\omega_{\nu})$ is the phonon self-energy. 
Note that within the conserving approximations, we must determine the phonon 
self-energy by differentiating the free-energy functional:
\begin{equation}
\Pi(i\omega_{\nu}) = \frac{-2}{T}\frac{ \delta \Phi }{ \delta D(i\omega_{\nu})}.
\end{equation}
With no vertex corrections, the phonon self-energy is simply given by the 
electron polarization, as depicted in Figure~\ref{fig:dself}(a). That is,
\begin{equation}
\Pi(i\omega_{\nu}) = U \pi^{(0)}(i\omega_{\nu}) .
\end{equation}
In this case, 
we use the contributions to the electron 
self-energy that are shown in Figure~\ref{fig:self}(c), whose explicit formula is 
given in Eq.~(\ref{eq:selfc}). Note that the term with the single polarization 
bubble is missing, as all orders of such `necklace' diagrams are included in the 
Fock term with the dressed phonon propagator. 

When we include vertex corrections, as well as dressed phonons, the phonon 
self-energy gains the extra term shown in Figure~\ref{fig:dself}(b). 
The full expression is given in the appendix 
[Eq.~(\ref{eq:dselfd})]. 
Note that a fully dressed phonon propagator is included in the phonon self-energy. 
As shown in Figure~\ref{fig:self}(d), the electron 
self-energy now has the expected extra term with a crossing of phonon lines. 
It is almost identical to the extra term 
in Figure~\ref{fig:self}(b), except of course, now the phonons are dressed. Again, 
the details of the formula can be found in the appendix.  

Our calculations all involve iteration of the Green's functions and self-energies, 
until a self-consistent solution is reached. 
We begin with the non-interacting Green's functions 
(set the self-energies to zero), and use it to calculate an initial estimate of the 
self-energies, according to Eqs.~(\ref{eq:selfa}-\ref{eq:dselfd}). The new 
self-energies are used to calculate updated Green's functions, according to 
Eqs.~(\ref{eq:greeng}) and~(\ref{eq:greend}). 
The procedure is iterated, so that at each step there is an updated self-energy, 
which includes a fraction of the 
previous self-energy, the exact fraction variable, 
dependent upon the progress of the iteration. 
We stop the process when the change in all the self-energies 
is less than one part in $10^{-10}$, which is typically after tens, 
but sometimes after 
hundreds of iteration steps. 

Superconductivity occurs below a critical temperature, $T_{c}$, where the normal 
state becomes unstable to fluctuations in the pairing potential (the Cooper 
instability). The instability shows itself as a divergence in the 
pairing susceptibility, $\raise2pt\hbox{$\chi$}_{m,n}$, which is given by:
\begin{equation}
\raise2pt\hbox{$\chi$}_{m,n} = \raise2pt\hbox{$\chi^{(0)}_{m}$}\delta_{m,n} - 
T \sum_{l}  \raise2pt\hbox{$\chi^{(0)}_{m}$} \Gamma_{m,l} \raise2pt\hbox{$\chi$}_{l,m} ,
\end{equation}
where $\Gamma_{m,n}$ is the irreducible vertex function, to be defined shortly. 
The bare susceptibility in the superconducting channel 
for momentum, ${\bf q}$, is defined as:
\begin{equation}
\raise2pt\hbox{$\chi^{(0)}_{m}$}({\bf q}) \equiv \frac{1}{N}\sum_{{\bf k}} 
G(i\omega_{m},{\bf k})G(-i\omega_{m},-{\bf k}+{\bf q}),
\end{equation}
which becomes in the local approximation (for the zero-momentum pair):
\begin{equation}
\raise2pt\hbox{$\chi^{(0)}_{m}$} = 
\frac{ \mbox{\large{\sl Im}\normalsize } 
[G(i\omega_{m}) ]}{\omega_{m}  Z(i\omega_{m}) },
\end{equation}
where 
\begin{equation}
\omega_{m} Z(i\omega_{m}) = \omega_{m} - \mbox{\large{\sl Im}\normalsize } 
[ \Sigma(i\omega_{m}) ].
\label{eq:zn}
\end{equation} 
The transition temperature, $T_{c}$, occurs when the 
largest eigenvalue of the matrix 
$-T \raise2pt\hbox{$\chi^{(0)}_{m}$} 
\Gamma_{m,l}$ passes through unity. 
We calculate the highest eigenvalue using the power method.

The irreducible vertex function, $\Gamma_{m,n}$ is given by 
the functional 
derivative of the off-diagonal self-energy, with respect to the off-diagonal 
Green's function:
\begin{equation} 
T \Gamma_{n,m} = \delta \phi(i\omega_{n}) /  \delta F(i\omega_{m}). \end{equation} 
As the pairing fluctuations lead to an instability of the normal state, 
$\Gamma_{n,m}$ must be calculated in the normal state
({\it i.e.} in the limit $F(i\omega_{n}) \mapsto 0$). 

Figure~\ref{fig:vert} shows the contributions to the irreducible vertex function for 
each of the four approximations. Each diagram is achieved by removal of one 
electron Green's function line, from a self-energy diagram in 
Figure~\ref{fig:self}. 
The algebraic expressions for each of the four sets of diagrams 
are given in the appendix. 

As we wish to uncover more than the phase diagram given by the different 
perturbation approximations, we move on to describe how we calculate other 
properties. In order to find the superconducting gap and 
thermodynamic quantities, calculations are required within the superconducting 
state, but a simple addition to the previous computations in the normal 
state allows us to calculate the isotope coefficient. 

The isotope coefficient, $\alpha$, describes how the critical temperature changes 
with the phonon mass, $M$. It is defined as:
\begin{equation}
\alpha = - \frac{ d \ln T_{c} }{d \ln M},
\end{equation}
so that $T_{c} \propto M^{-\alpha}$. 
The weak-coupling limit of BCS theory, and Migdal-Eliashberg theory with no 
Coulomb repulsion, predict $\alpha = 0.5$, which corresponds to 
$T_{c} \propto 1/\sqrt{M}$. 
The phonon frequency changes with mass, according to 
$\Omega \propto  1/\sqrt{M} $, so both the product, $M\Omega^{2}$, and the 
interaction energy, $U$, remain constant. Hence we 
calculate the isotope coefficient simply by changing the phonon frequency by $1$\% 
(corresponding to a typical mass change of $2$\% between isotopes) 
and comparing the change in critical temperature. To be precise, 
we compute $\alpha$ by:
\begin{equation}
\alpha = 0.5 \cdot \frac{T^{(new)}_{c} - T^{(old)}_{c} }{T^{(old)}_{c} }
\cdot \frac{ \Omega^{(old)} }{\Omega^{(new)} -  \Omega^{(old)} },
\end{equation}
so the BCS result is achieved if $T_{c} \propto \Omega$.

According to standard methods~\cite{Carb,Scal,Alle}, 
the energy gap in the superconducting state, $\Delta$, requires a self-consistent 
calculation within the superconducting state. Note, 
the order parameter on the imaginary axis is related to the off-diagonal 
self-energy through:
\begin{equation}
\Delta(i\omega_{n}) = \phi(i\omega_{n}) / Z(i\omega_{n}),
\end{equation}
where $Z(i\omega_{n})$ is the mass-enhancement parameter, calculated from the 
electronic self-energy, $\Sigma (i\omega_{n})$, as given in Eq.~(\ref{eq:zn}). 
The gap itself is found from 
the order parameter on the real axis, at the point where 
$\mbox{\large{\sl Re}\normalsize } \left[ \Delta (\omega) \right] = \omega$. 
We carry out a Pad\'{e} analytic continuation~\cite{Sere,Leav} 
to obtain the order parameter on the real axis, 
and hence the value of the gap. 

The free energy can be found from the 
formula~\cite{Elia2,BaSt,Lutt}:
\begin{eqnarray}
F & = & -2T \sum_{n} \left\{
\frac{1}{2} \ln \left[-  1 / \det \underline{\underline{G}}(i\omega_{n}) \right] 
+ \frac{1}{2} {\rm Tr} \underline{\underline{\Sigma}} (i\omega_{n})
\underline{\underline{G}}(i\omega_{n}) \right\} 
\\* \nonumber
& & + \frac{T}{2} \sum_{\nu} \left\{ 
 \ln \left[ - 1 / D(i\omega_{\nu}) \right] + 
\Pi(i\omega_{\nu}) D(i\omega_{\nu}) \right\} 
\, + \,\Phi \, + \mu (n-1).
\end{eqnarray}
The free-energy functional, $\Phi$, is the skeleton-diagram expansion
whose differential with 
respect to the electron (or phonon) Green's function gives the electron (or phonon) 
self-energy [see Figure~\ref{fig:free} 
and Eqs.~(\ref{eq:freea}), (\ref{eq:freeb}),
(\ref{eq:freec}), and (\ref{eq:freed})]. 
We are interested in the free-energy difference between normal and superconducting 
states at fixed electron filling, $n$, in which case the first, Hartree, 
term in $\Phi$ is neglected as it is a constant, $Un^{2}/2$. The final term, 
$\mu (n-1)$, can not be neglected, because the chemical potential can differ 
considerably between the superconducting and normal state when one includes the
effects of non-constant density of states. 

We calculate the thermodynamic critical field in the superconducting state, $H_{c}$, 
from the free-energy difference between the superconducting and normal state, 
according to the formula \mbox{$F_{S}-F_{N} = -\mu_{0} H_{c}^{2}$}. 
The thermodynamic field 
varies with temperature in an almost quadratic manner, so that calculation of the 
deviation function, which is defined as the difference between $H_{c}(T)$ and the 
quadratic form, $H_{c}(0)\left[ 1 - (T/T_{c})^{2}\right]$ gives a sensitive test of 
changes in thermodynamic quantities.

In the calculations which include a Coulomb repulsion term $U_{C}$, we make the 
standard simplification~\cite{Mars,Carb,Scal,Alle} 
of only including its effects on the off-diagonal self-energy. This simplification is 
valid, as the normal-state Green's functions in reality include the Coulomb 
repulsion effects, and these change by very little for the diagonal part of the 
Green's function when superconducting order is present. As our model does not 
include the effects of $U_{C}$ on the diagonal Green's functions, it is not strictly 
the solution of a simple Hamiltonian with a $U_{C}$ term included~\cite{Free5}. 
However, the simplification does allow us to compare the effects of including a 
Coulomb repulsion versus adding vertex corrections, on the 
superconducting properties and transition temperature from a similar normal 
state and it is precisely the method employed in analyzing experimental data
on real materials. 
With these comments understood, the only changes to the calculations that are 
necessary with the inclusion of a Coulomb term, are that both the off-diagonal 
self-energy, $\phi(i\omega_{n})$, and the irreducible vertex function, 
$\Gamma_{n,m}$ gain an extra term:
\begin{eqnarray}
\phi(i\omega_{n}) &  \mapsto & \phi(i\omega_{n}) + U_{C}T\sum_{m} F(i\omega_{m}),
\\*
\Gamma_{n,m} & \mapsto & \Gamma_{n,m} + U_{C}.
\end{eqnarray}
In the computational calculations, the Matsubara frequency sum is cut off 
after a constant number, $N_{c}$, of terms. 
This leads to a renormalization which reduces the 
Coulomb term~\cite{Mars,Bogo} 
to a pseudopotential, $U_{C}^{*}$, given by:
\begin{equation}
U_{C}^{*} =  U_{C} \left/ \left\{ 1 - 2TU_{C} \sum_{N_{c}+1}^{\infty} 
\frac{\mbox{\large{\sl Im}\normalsize } 
\left[ G(i\omega_{m}) \right] }{\omega_{m}} \right\} \right. ,
\end{equation}
where the diagonal Green's function, $G(i\omega_{m})$, is at high frequency, 
where the self-energies may be neglected, 
but must include the self-consistent chemical potential, $\mu$. 
In our calculations, with 
the frequency cut-off on the scale of the bandwidth such that 256 or 512 
Matsubara frequencies are used, there is a very small (typically $1\%$ or $2\%$) 
reduction in $U_{C}$. This is different from the conventional approach in
real materials because here our energy cutoff is governed by the electronic
bandwidth, not some multiple of the maximum phonon frequency.
To make contact with the standard formalism, we define a 
dimensionless pseudopotential, $\mu^{*}_{C} = \rho^{(0)}(\mu) \cdot U_{C}^{*}$, 
where $\rho^{(0)}(\mu)$ is the non-interacting electronic density of states at the 
chemical potential. In calculations at different temperatures, $U_{C}$ is kept 
fixed, so that $\mu^{*}_{C}$ varies to a small extent. 
 
Finally, we wish to make clear how $\lambda$, the measure 
of the electron-phonon coupling strength, is defined in our work. A precise definition 
is required, because different methods of calculating $\lambda$ lead to different 
results away from the weak-coupling limit, especially when the phonons are dressed. 
Here, $\lambda$ is given by:
\begin{equation}
\lambda = \rho^{(0)}(\mu) \cdot U \cdot D(0)
\end{equation}
where $D(0)$ is the zero-frequency component of the {\it dressed} 
phonon propagator, which can be significantly different from that of the `bare' 
propagator, $D^{(0)}(0)$. This value of $\lambda$ is usually different from the 
electronic mass 
enhancement parameter, {\it i.e.} $\lambda \neq Z(0) - 1$, as the two are only 
equal in the weak-coupling limit and with $\Omega \mapsto 0$. The definition of 
$\lambda$ is identical to that commonly calculated from the electron-phonon spectral 
density~\cite{Carb}, $\alpha^{2}F(\omega)$, namely:
\begin{equation}
\lambda = 2\int^{\infty}_{0} \frac{\alpha^{2}F(\omega) d\omega}{\omega}
\end{equation}
where  
\begin{equation}
\alpha^{2}F(\omega) = \rho^{(0)}(\mu) \cdot |U| \cdot 
 \frac{1}{\pi} \mbox{\large{\sl Im}\normalsize } \left[ D(\omega) \right]
\end{equation}
The real-axis form of the phonon propagator, $D(\omega)$, is calculated from its 
imaginary-axis values, $D(i\omega_{\nu})$, by a Pad\'{e} analytic continuation.

\section{Results}

In choosing the parameters used to carry out the calculations, a number of criterion 
had to be satisfied. Firstly, we wished to operate outside the weak-coupling 
regime, so that vertex corrections would not be negligible. The phonon frequency 
needed to be large compared to conventional low-temperature
superconductors, but not so 
large that it gave no realistic point of contact with those superconductors 
mentioned in the introduction. So we choose  $\Omega=t$, equal to one-twelfth 
of the bandwidth. 
We had to ensure the electron-phonon coupling strength was not 
so strong that the ground state would contain bipolarons~\cite{Alex}, making the 
perturbation expansion about a Fermi liquid state invalid. A maximum coupling of 
$U=-2t$, hence a bare $\lambda < 0.5$ ensured this. Finally, in order to calculate 
properties in an achievable time, while ensuring the imaginary frequency cut-offs 
were at energies larger than the band-width, the temperature of the calculations 
could not be too small, hence $T>2\times 10^{-3}t$. The last condition meant that results 
within the superconducting state were best carried out for as large $U$ and $\Omega$ 
as possible, so that $T_{c}$ would be high. Hence, calculations near the 
band-edges, where the density of states was low, prove to be inaccurate, due to the 
very low transition temperatures there. 

\begin{figure}
  \centerline{\psfig{figure=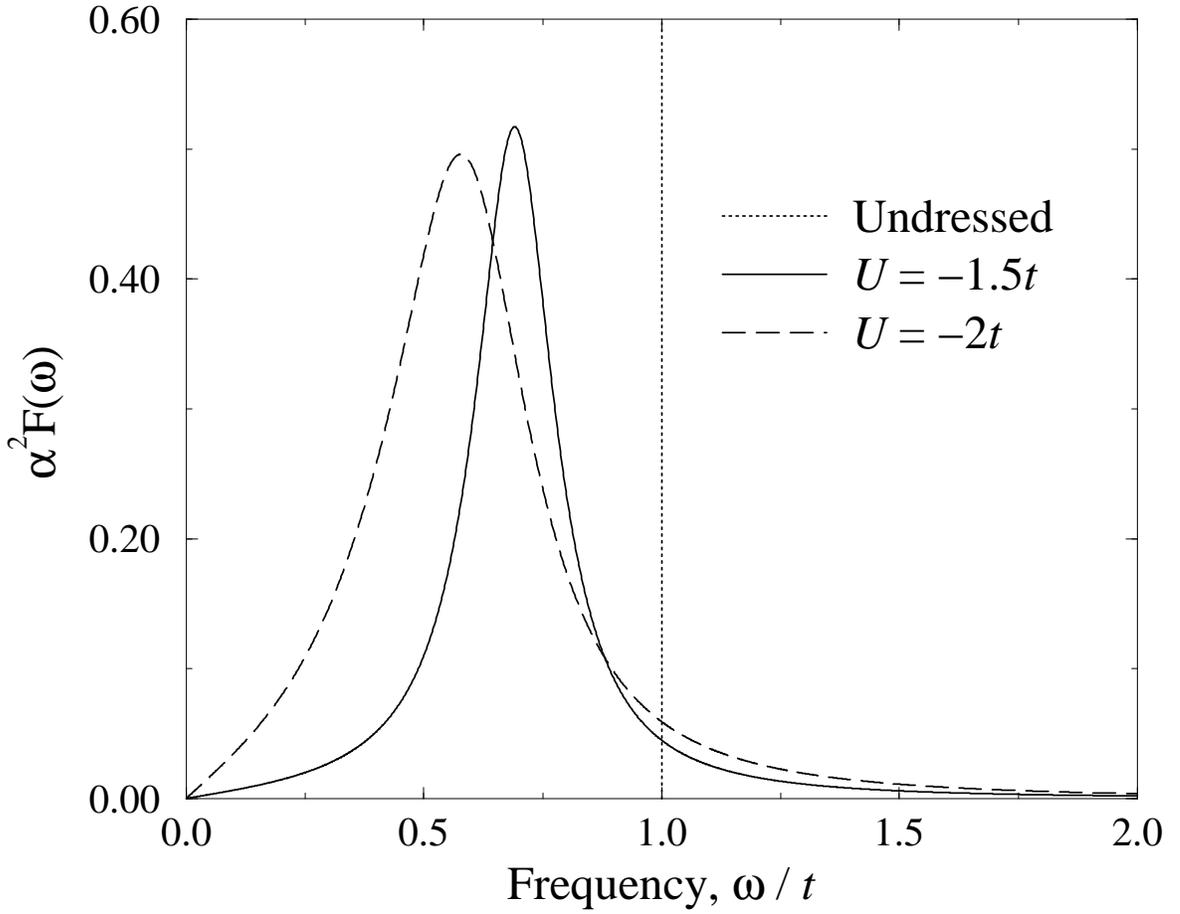,width=6.0in}}
\caption{$\alpha^{2}F(\omega)$, calculated from the dressed phonon propagator on the 
real frequency axis for $n=1$ and $\Omega=t$. Increased coupling leads to a greater 
downwards shift in frequency from the initial delta-function at $\omega=t$.}
\label{fig:dofw}
\end{figure}

Our first result is 
that dressing the `bare' phonon propagator leads to considerable 
renormalization effects. 
To be specific, the 
value of $\lambda$ doubles from it's `bare' value of $\lambda = 0.21$, to 
$\lambda \approx 0.4$ after dressing the phonons, using 
parameters 
$\Omega = t$ and $U=-1.5t$, near half-filling. 
Moreover, at the increased interaction strength 
of $U=-2t$, 
$\lambda$ is enhanced by a factor of {\it three} from the value of 
$\lambda=0.28$ for undressed phonons to the 
dressed value of $\lambda \approx 0.9$. Such an enhancement indicates that the 
perturbation expansion would be 
inaccurate at bare coupling strengths lower than might be 
naively expected. Figure~\ref{fig:dofw} shows how $\alpha^{2}F(\omega)$ is altered 
from its `bare' value, 
a delta-function situated at $\omega = \Omega= t$, 
when it is dressed. Note that there is both 
a shift to lower frequencies as well as a broadening of the spectrum. 
The shift to lower 
frequencies shows that a Holstein model with `bare' phonon frequencies of near 
$10\%$ of the bandwidth can be required to lead to dressed phonon frequencies at 
approximately $5\%$ of the bandwidth. Hence the exact phonon self-energy used 
is an important factor when modeling electron-phonon systems, 
near the crossover between the weak-coupling and strong-coupling regimes.

\begin{figure}
  \centerline{\psfig{figure=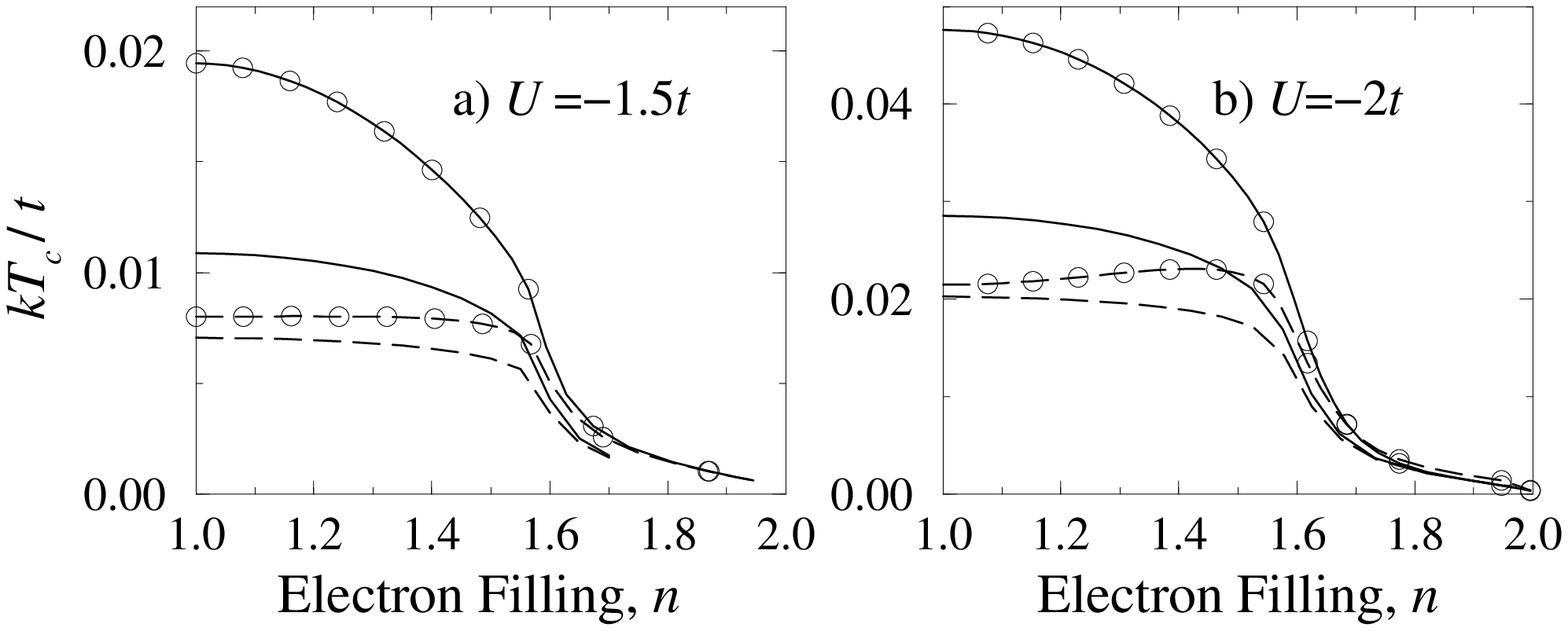,width=6.5in}}
\caption{Transition temperature as a function of filling. The dashed lines indicate 
vertex corrections are included, while solid lines are without them. The lines with 
circles have dressed phonons. Both diagrams show an enhanced $T_{c}$ by dressing 
the phonons, while vertex corrections reduce $T_{c}$ near half-filling. 
In all cases, $T_{c}$ falls rapidly above a filling of $n=1.55$, 
where the density of states drops rapidly.}
\label{fig:Tc}
\end{figure}

The preceding paragraph explains some of the dramatic differences between the 
critical temperature ($T_{c}$) 
values for the different approximations, shown in Figure~\ref{fig:Tc}. 
In particular, the 
$T_{c}$ for dressed phonons is markedly higher 
than that for bare phonons, which is 
to be expected as $\lambda$ is also higher. Note that the transition temperatures fall 
rapidly with increasing filling, above a filling of about $n=1.5$, 
as the electronic density of states drops significantly in this region. Vertex 
corrections seriously reduce $T_{c}$ near half-filling ($n=1$).
The two curves with dressed phonons in Figure~\ref{fig:Tc} show a greater disparity 
than the two curves with `bare' phonons in the same figure, which means that 
dressing the phonons, which increases the effective coupling, enhances the effect 
of vertex corrections. Although it is hard to distinguish the curves due to the low 
$T_{c}$ near the band edge, above a filling of $n=1.75$ the vertex corrections do lead 
to an enhancement of $T_{c}$. This result is in agreement with previous 
work~\cite{Free2,Nico}. 
Note that all results show particle-hole symmetry, that is, they are symmetric about 
half-filling. The figures just show half of the band ($n>1$), neglecting a mirror 
image below half-filling.

It is clear that vertex corrections do change $T_{c}$ by a considerable amount, 
but for any experimental measurement, where the microscopic parameters 
are not known, a Coulomb pseudopotential, $\mu^{*}_{C}$, can always 
be fitted to give the same $T_{c}$ as vertex-corrections. Hence we continue 
with other results, to see where vertex corrections can not simply be mimicked by 
an appropriate $\mu^{*}_{C}$, which would cause the effects of vertex corrections 
to be unobservable. So we fit a $\mu^{*}_{C}$ to give the same value of $T_{c}$ 
as vertex corrections, and go on to change the unobservable electron-phonon 
coupling strength, $U$, to give the same $\lambda$ as vertex corrections, when 
phonons are dressed. Hence the effects of vertex corrections can show up as 
discrepancies over a range of quantities compared to the values obtained 
with a fitted $\mu^{*}_{C}$ and $\lambda$.

\begin{figure}
  \centerline{\psfig{figure=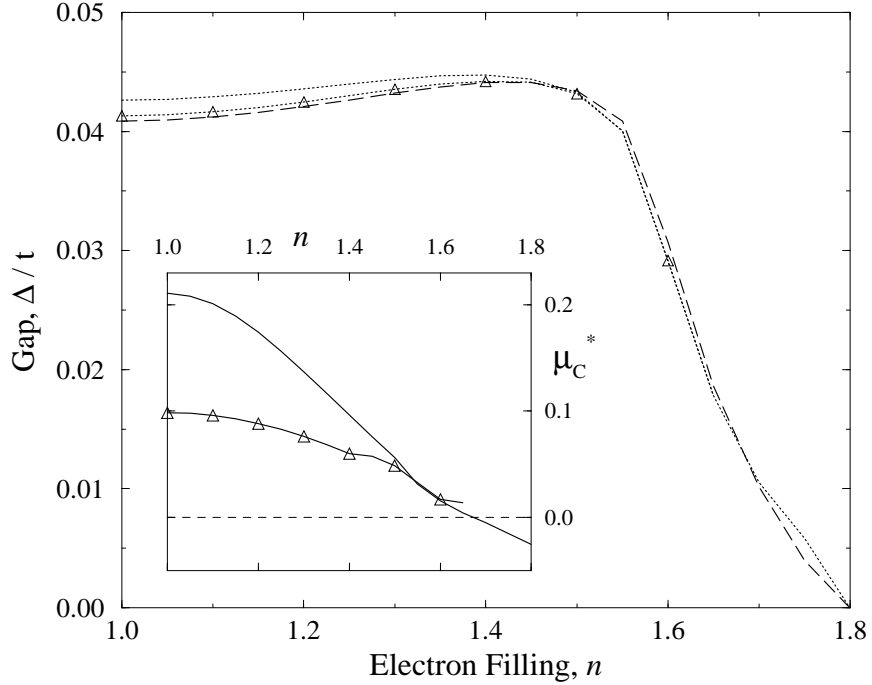,width=4.5in}}

\caption{A comparison between the effects of vertex corrections and a Coulomb 
pseudopotential, $\mu^{*}_{C}$ on the superconducting gap. The calculations are 
with dressed phonons, with $U=-2t$ and $\Omega=t$. The dashed curve is with vertex 
corrections, while the dotted curves include a $\mu^{*}_{C}$, whose value changes 
with filling as shown in the inset, to ensure the two corresponding $T_{c}$ curves are 
exactly the same. The dotted curve with triangles indicates a fit to the same 
$\lambda$ by adjusting $U$ as well as $\mu^{*}_{C}$.}
\label{fig:ucvert}
\end{figure}

Figure~\ref{fig:ucvert} is a direct comparison between the effects of vertex 
corrections and a Coulomb pseudopotential, $\mu^{*}_{C}$, 
on the value of the gap parameter. Our first method is to 
fix the bare electron-phonon coupling and to 
adjust the value of $U_{C}$ 
and hence $\mu^{*}_{C}$ in a calculation without vertex corrections, until the 
same $T_{c}$ is reached as found in the calculation with vertex corrections 
(shown in Figure~\ref{fig:Tc}(b)). 
$U_{C}$ is then used unchanged, to calculate other properties such as the gap 
parameter. The Coulomb pseudopotential
$\mu^{*}_{C}$ varies with electron filling at $T_{c}$ as shown in the 
inset. 
Notice that an artificial value of 
$\mu^{*}_{C}<0$ is required when $n>1.7$, as vertex corrections enhance $T_{c}$ 
in this region. Near half-filling, where $\mu^{*}_{C}$ is positive, 
and reduces $T_{c}$ like vertex corrections, the gap parameter is reduced by a 
smaller amount. Hence a Coulomb repulsion leads to slightly higher gap ratio than 
vertex corrections. 

The second curve, with a lower value of $\mu^{*}_{C}$, 
is an alternate approach, where both the dressed value of $\lambda$ and $T_{c}$ 
are fitted to the results with vertex corrections. A simple fit to $T_{c}$ 
with a fixed bare coupling leads 
to a higher $\lambda$ with a Coulomb repulsion than with vertex corrections, 
because $\lambda$ is determined through the dressed phonon propagator. 
As a conventional analysis would fit $\lambda$ to the experimental data, as well as 
$\mu^{*}_{C}$, this second method follows the spirit of our paper by 
trying to fit conventional theory to the vertex-corrected results. 
In order to give the same values of $\lambda$, 
the electron-phonon interaction energy had to be varied, and in fact reduced by 10\% 
at half-filling. The result is a lower value of the gap ratio than with only 
$\mu^{*}_{C}$ fitted, but still a slightly larger value than with vertex corrections 
alone. 

\begin{figure}
  \centerline{\psfig{figure=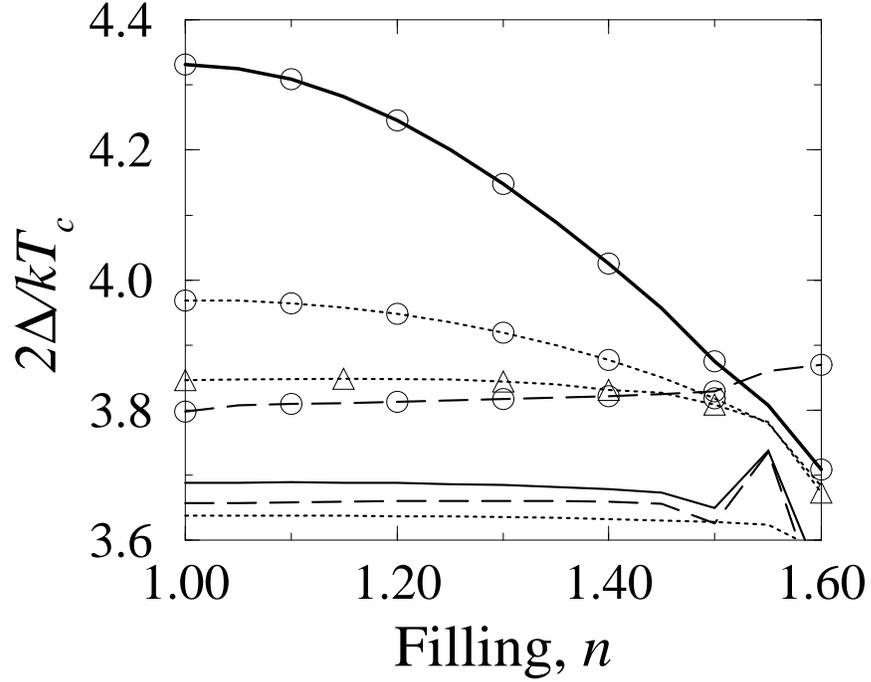,width=4.5in}}

\caption{Gap ratio, $2\Delta/ kT_{c}$, as a function of filling, $n$. Dressed 
phonons (with circles) exhibit strong coupling behavior, by the increased 
gap ratio. Vertex corrections (dashed lines) reduce the effective coupling 
strength in the center of the band, hence the gap ratio is lower in this region. 
The dotted lines indicate how a Coulomb pseudopotential alters the gap ratio, 
(when $T_{c}$ is matched to $T_{c}$ with vertex corrections, and where triangles 
indicate that $\lambda$ is also matched, by adjusting $U$). With dressed phonons, 
the Coulomb repulsion clearly has less of an effect than do vertex corrections, 
but this is not the case with bare phonons.}
\label{fig:rat}
\end{figure}

Although the magnitude of the gap 
varies considerably, depending upon the approximation used, Figure~\ref{fig:rat} 
shows that the gap ratio, $2\Delta /kT_{c}$, varies less markedly. The gap ratio is 
greater than $4$ in the case of dressed phonons without vertex corrections, which is 
typical of the strong coupling regime ($\lambda > 0.5$). 
Note that when the phonons are 
`bare', so the coupling is less strong, the inclusion of vertex corrections, while 
strongly reducing $T_{c}$ and $\Delta$ individually, has little effect on the ratio, 
$2\Delta/ kT_{c}$. 

The isotope coefficient, $\alpha$, has a value of $0.5$ in the simplest, BCS, 
approximation, and in Migdal-Eliashberg theory with no Coulomb repulsion. The reason 
is that the phonon frequency provides the only cut-off for the coupling between 
different states, and phonon frequencies are proportional to $M^{-0.5}$, where $M$ is 
the ionic mass. 
Inclusion of a frequency-independent Coulomb repulsion, $U_{C}$,  
leads to a 
reduction in $\alpha$, as does a finite bandwidth. The reduction in $\alpha$, 
indicates that the increase 
in phonon frequency is less effective at increasing $T_{c}$ than otherwise. 
Higher-frequency phonons reduce the retardation in the electron-electron attraction, so 
the Coulomb repulsion between electrons is less shielded. The finite bandwidth 
means that the number of states coupled together includes a factor independent of 
phonon frequency, so $T_{c}$ does not increase with $\Omega$ as it might if there 
were an infinite number of electron states extending through all energies. 

It is known~\cite{Garl,Mars} that 
Migdal-Eliashberg theory with a finite bandwidth and including a Coulomb repulsion, 
leads to the allowed range of 
values for the isotope coefficient, $ 0 \leq \alpha \leq 0.5$. 
Including a non-constant 
density of states~\cite{Scha,CaNi} can in principle lead to any positive 
value of $\alpha$. The reason 
being that $T_{c}$ increases because extra electron states near the chemical potential 
are able to couple together when the phonon frequency increases. If the density of 
states increases significantly in the region of energy where new states are coupled 
together, the increase in $T_{c}$ is much higher than would otherwise be expected, and 
$\alpha$ can be large, even greater than $0.5$. 
The corollary is that if the density of states 
decreases away from the chemical potential, $\alpha$ also decreases, but never to less 
than zero, as $T_{c}$ does not go down when the number of 
states coupled together goes up. 

\begin{figure}
  \centerline{\psfig{figure=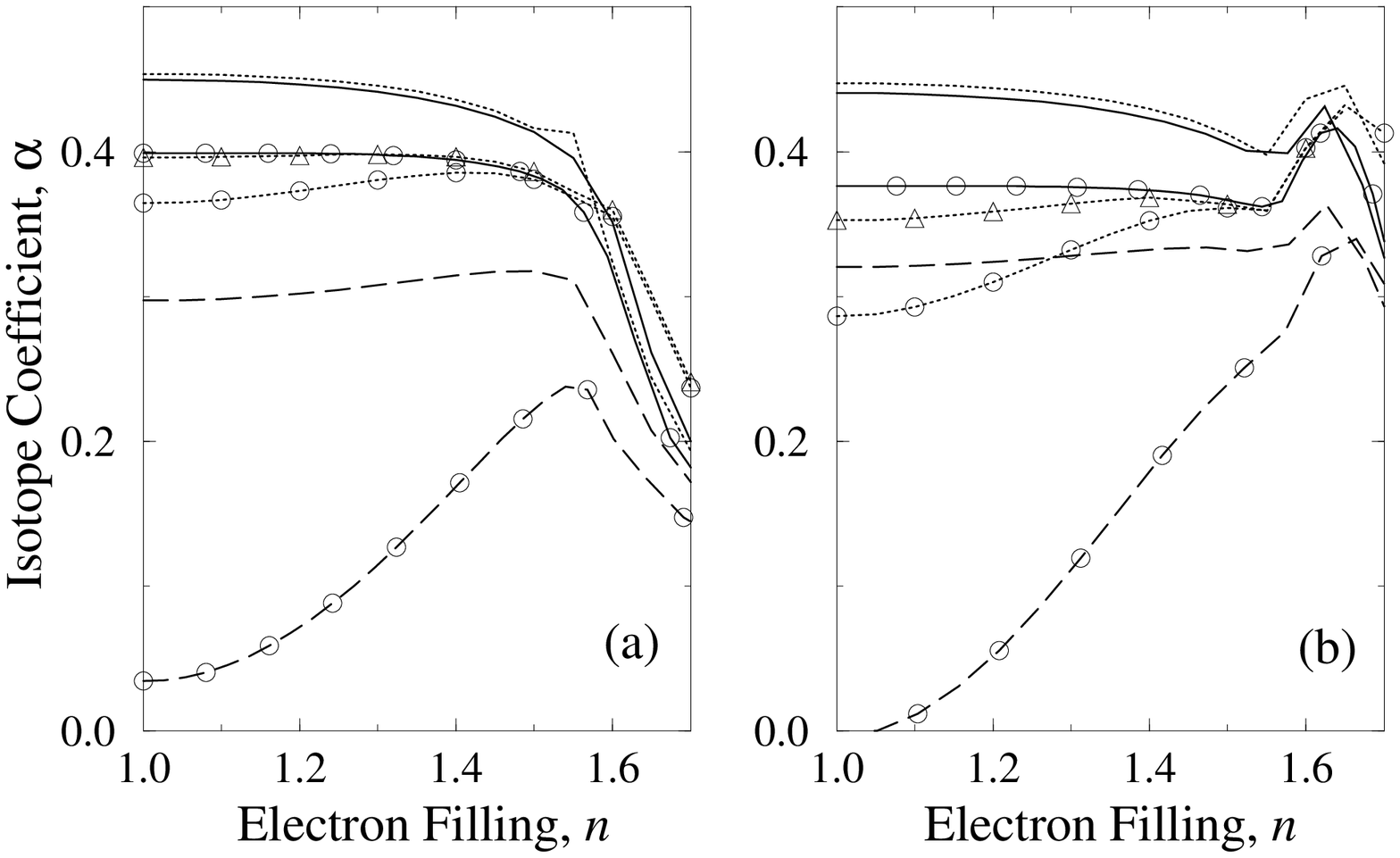,width=6.0in}}
\caption{The isotope coefficient, $\alpha$, 
plotted against electron filling, $n$, for (a) $U=-1.5t$ and (b) $U=-2t$.
Vertex corrections, indicated by dashed lines, reduce $\alpha$. As does 
dressing the phonons, seen by the lines with circles. Figure (b) exhibits 
the unusual feature of $\alpha < 0$ for dressed phonons, with vertex 
corrections included. The dotted curve in (b) is the result with the Coulomb 
repulsion, $\mu^{*}_{C}$, shown in the inset in the previous figure.}
\label{fig:iso}
\end{figure}

Figure~\ref{fig:iso} shows that the inclusion of 
vertex corrections in the calculations 
with dressed phonons not only reduces $\alpha$, but can in fact 
lead to negative values. Indeed, the strongest reduction in $\alpha$ by vertex 
corrections occurs near half-filling, and at strong coupling, where $T_{c}$ is 
comparatively large. 
By comparison, in all cases the Coulomb pseudopotential, 
which gives 
the same reduction in $T_{c}$ as vertex corrections, leads to a much smaller 
reduction in $\alpha$. In Migdal-Eliashberg theory, 
a very small value of $\alpha$ requires a very low $T_{c}$. 

Hence, any observation of isotope effects which 
have $\alpha < 0$, or a small $\alpha$ with moderate to high $T_{c}$, 
implies that either vertex corrections are involved or some
other mechanism outside of Migdal-Eliashberg theory is important. 
Paramagnetic impurities~\cite{Para,Bill}, proximity effects~\cite{Bill2}, 
anharmonicity~\cite{Drec,Galb}, and 
an isotopic dependence of the electron density in the conduction band~\cite{Bill,Bill2}, 
can also lead to a low or negative $\alpha$ without requiring a low $T_{c}$. One or 
more of these effects may be important, in those materials with anomalously low 
values of $\alpha$~\cite{Park,Ur,PdH,LSCO,LSCO1,LSCO2,C60a,C60b}, 
but vertex corrections should also be considered.

An important effect of dressing the phonons is that a small increase in 
the bare phonon frequency does not just 
result in a constant shift of $\alpha^{2}F(\omega)$, through a rescaling of the 
frequency variable. When calculating the isotope effect, it is common to assume 
that a mass substitution simply rescales the frequency, otherwise maintaining 
the same form of $\alpha^{2}F(\omega)$. 
However, the phonon self-energy is not independent of frequency, so 
the magnitude of  $\alpha^{2}F(\omega)$ at its peak, which is inversely 
proportional to the imaginary part of the self-energy at that frequency, does not 
remain constant. In fact, the peak height is reduced by an increase in peak 
frequency, resulting in a slight reduction in $\lambda$. Hence the 
increase in $T_{c}$ with bare frequency is less than otherwise expected, 
reducing the isotope 
coefficient for dressed phonons.  

Interestingly, when the phonons are undressed, the 
Coulomb repulsion leads to a very small increase in $\alpha$. This arises, because 
the term with a single polarization bubble (the first order term coming from 
dressed phonons) is present. The term acts to reduce $\alpha$, but is less 
significant at the lower transition temperatures caused by the 
Coulomb pseudopotential. Meanwhile, the increase in $\mu^{*}$ with temperature, 
which acts to reduce $\alpha$ is a much smaller effect.

\begin{figure}
  \centerline{\psfig{figure=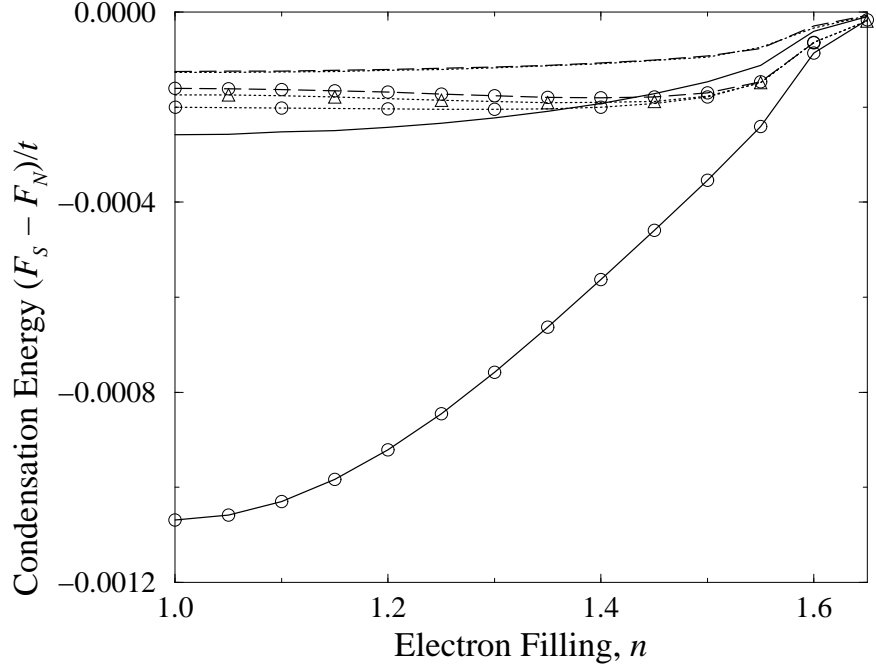,width=4.5in}}

\caption{Condensation energy, $F_{S}-F_{N}$. Circles indicate the phonons are 
   dressed, which leads to a greater condensation energy. The dotted lines 
   indicate dressed phonons with a Coulomb pseudopotential, to mimic the 
   vertex-corrected $T_{c}$, the triangles indicating that $U$ is also 
   adjusted to mimic the $\lambda$ obtained with vertex corrections. 
   The dashed, vertex-corrected curve shows smaller 
   condensation energy for $n<1.55$, though the difference can not be seen with 
   bare phonons.}
\label{fig:endiff}
\end{figure}

The free-energy difference, $\Delta F=F_{S} - F_{N}$, 
is plotted as a function of filling, $n$, 
in Figure~\ref{fig:endiff}. In the simplest picture, 
true in the weak-coupling limit, one expects the magnitude 
of the free-energy 
difference to be approximately equal to $\rho(\mu) \Delta^{2}/2$, 
representing a number of 
states, $\rho(\mu)\Delta$, each shifted by an average energy of order $\Delta/2$. 
Although, with $\rho(\mu)$ given by $Z(0)\rho^{(0)}(\mu)$, the weak-coupling result 
predicts too high a condensation energy~\cite{Scal}, it does explain the 
qualitative changes between the different curves of Figure~\ref{fig:endiff}. 

In fact, the dimensionless quantity, $\gamma T_{c}^{2}/(8\pi\Delta F)$
[the Sommerfeld constant, $\gamma = 2\pi^{2}k_{B}^{2}\rho^{(0)}(\mu)Z(0)/3$] 
changes little for these 
curves. At half-filling, with dressed phonons, the value is reduced from the BCS 
constant result of 0.168 to a strong-coupling value of 0.137. The value with vertex 
corrections is 0.157, while with the Coulomb pseudopotential it is 0.149. Hence, 
as with the gap ratio, near half-filling, 
vertex corrections cause quantities to be closer to the 
weak-coupling values than does a Coulomb pseudopotential fitted for the same 
$T_{c}$. The values at a filling of $n=1.6$ are all closer to the 
weak-coupling limit, as expected when the density of states falls. The result is 
0.163 for dressed phonons without vertex corrections, changing little to 0.165 
with a Coulomb pseudopotential and 0.158 with vertex corrections.

It is worthwhile pointing out that the difference in thermodynamic potentials, 
which is usually calculated as an approximation to the free-energy difference, 
leads to very different results away from half-filling. 
The approximation is based on the assumption that the 
chemical potential changes little 
between normal and superconducting states, but this is not necessarily the case 
when there is a non-constant 
density of states. In fact, there is a particularly large shift from the 
normal-state chemical potential, $\mu_{N}$, to that in the superconducting state,  
$\mu_{S}$, if $\mu_{N}$ lies 
near the van Hove singularity, where the non-interacting 
density of states is falling precipitously. 
This can be understood, by considering how states above and below the 
normal-state chemical potential, $\mu_{N}$, couple together and create an energy gap. 
When the density of 
states is falling rapidly with increasing energy, electronic states from a larger 
energy region above $\mu_{N}$ couple to those in a small region below $\mu_{N}$. The 
resulting energy gap is skewed up in energy, so the chemical potential in the 
superconducting state, $\mu_{S}$, which sits in the middle of the gap, becomes 
greater than $\mu_{N}$. In the mirrored example 
below half-filling, near $n=0.4$, $\mu_{S}$ is also 
pushed away from half-filling, so we find $\mu_{S} < \mu_{N}$ there.

\begin{figure}
  \centerline{\psfig{figure=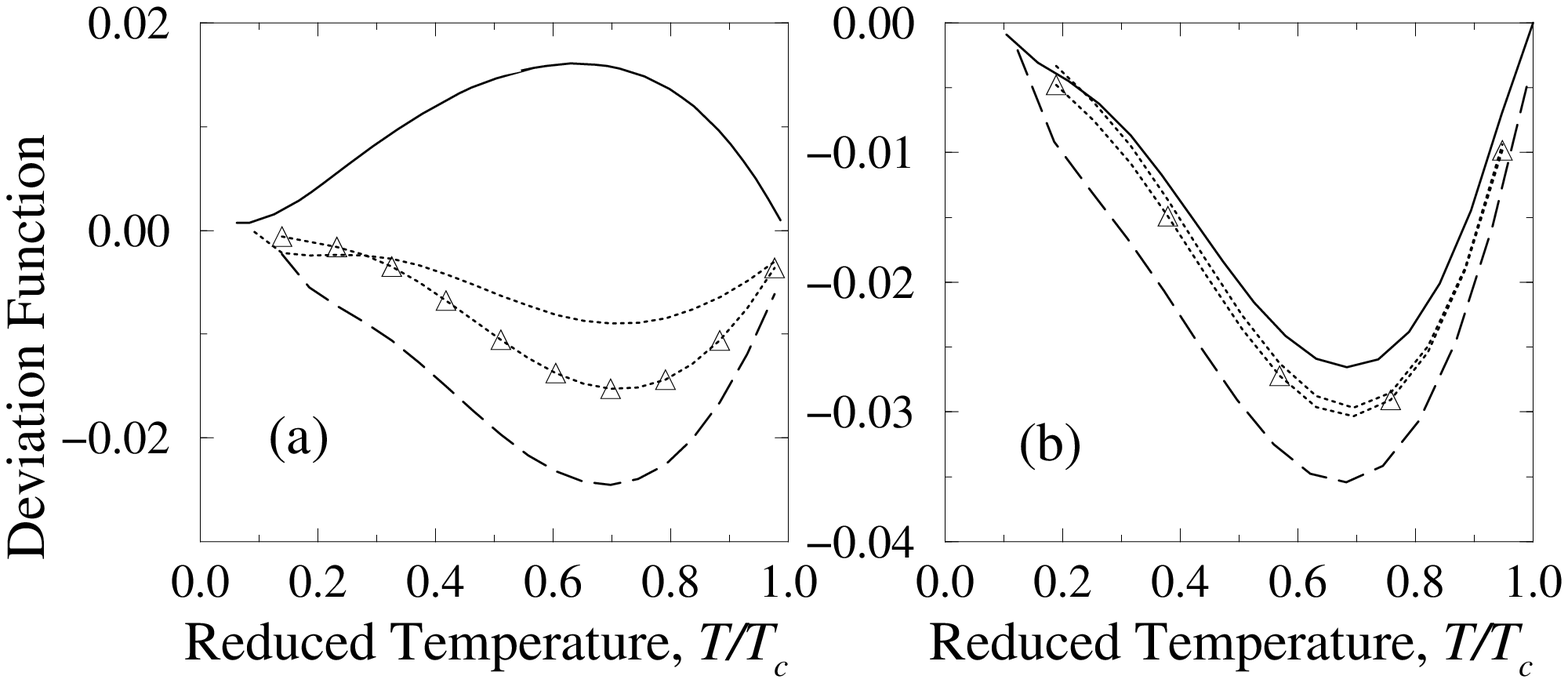,width=6.0in}}

\caption{Critical-field deviation function, $H_{c}(T)/H_{c}(0) - [1-(T/T_{c})^{2}]$. 
Solid lines are without, dashed lines are with vertex corrections. The dotted lines 
with and without triangles 
include the Coulomb pseudopotentials, $\mu^{*}_{C}$ of 
Figure~\ref{fig:ucvert} inset. All results are 
for dressed phonons. (a) is at half-filling, $n=1$ while 
(b) is at a filling of $n=1.6$. Note the shift in scale 
for the second graph.} 
\label{fig:dev}
\end{figure}

The thermodynamic critical field, $H_{c}$, 
is effectively the square root of the free-energy 
difference, so shows qualitatively the same effects. 
$H_{c}$ is known~\cite{Carb} to vary with temperature 
in a manner close to the behavior 
$H_{c}(T) = H_{c}(0)\left[ 1 - (T/T_{c})^{2} \right]$ 
which corresponds to the two-fluid model. 
The deviation function, plotted in Figure~\ref{fig:dev}, is the 
difference between the reduced 
critical field, $H_{c}(T)/H_{c}(0)$, and the quadratic fit, 
$1 - (T/T_{c})^{2}$. 

The curve without vertex corrections at half-filling shows  
a small positive deviation, with a maximum of $0.02$, 
typical of strong-coupling superconductors. 
Interestingly, when vertex corrections are included, the 
deviation function at half-filling becomes negative, showing a minimum of 
about $-0.03$, which is more typical of weak-coupling superconductors (BCS theory 
predicts a minimum of $-0.037$). This result fits in with those for other properties, 
demonstrating that vertex corrections reduce the effective coupling strength, 
near half-filling. A Coulomb pseudopotential, with the same power to reduce $T_{c}$ 
as vertex corrections, also decreases the deviation function, but to a lesser extent 
than vertex corrections do so. This is still true when the value of $\lambda$ is also 
fitted, by altering $U$ as necessary. The calculations away from the band-center, 
at $n=1.6$, lead to a negative deviation for all approximations. The reduced 
density of states at this filling leads to weak-coupling behavior. 

Other thermodynamic quantities, which can be derived from the free-energy data, 
are affected in similar ways. That is, vertex corrections reduce the effective 
coupling strength, to a greater extent than does a Coulomb repulsion giving the 
same $T_{c}$. 
For example, 
vertex corrections reduce the specific heat jump, $\Delta C$ at $T_{c}$, 
as does a Coulomb repulsion to a lesser extent. The following results are obtained 
by a numerical differentiation, so are not completely accurate in themselves, 
(perhaps only to 10\%) but as much of the error is systematic, the trends are 
reliable. 
At half-filling, the dimensionless quantity, $\Delta C/ \gamma T_{c}$ 
is reduced from the strong-coupling value of 2.44, to 1.63 with vertex corrections and 
1.88 with a Coulomb pseudopotential (the BCS result is 1.43). Again, 
notice the typical result that near half-filling, vertex-corrections lead to a 
weaker-coupling result than does inclusion of a Coulomb pseudopotential. 
At an electron filling of $n=1.6$, the results indicate less strong coupling, 
giving $\Delta C/ \gamma T_{c} = 1.66$ for dressed phonons, with the value reduced 
to 1.53 by vertex corrections, and to 1.44 by a Coulomb repulsion.

\section{Conclusions}

We have completed a numerical investigation of the effects of vertex corrections, 
dressing phonons and a non-constant density of states on the physical properties of 
strong-coupling superconductors. We solved the Holstein model, using four distinct 
perturbation theories, within conserving approximations. 

The necessity of incorporating a realistic phonon self-energy 
is of considerable importance to those working with model Hamiltonians. 
The use of dressed phonons in the Holstein model, leads to a large 
renormalization of the parameters - in particular, the value of $\lambda$ can 
be enhanced by a factor of $3$, when its `bare' value of $0.28$ would suggest the 
system is in the weak-coupling regime. Such an enhancement of $\lambda$ 
reveals itself in increased $T_{c}$, $\alpha$, $\Delta$, $H_{c}$, and a gap ratio 
($2\Delta/kT_{c}$) greater than 4. The real-frequency data shows that the Einstein 
spectrum (a delta-function at $\Omega$) is both broadened and peaked at a lower 
frequency, when the bare, Einstein phonons are dressed.

The non-constant density of states affects Migdal-Eliashberg results in both expected 
and unexpected ways. Firstly, all quantities which depend on the density of states 
as a parameter within Migdal-Eliashberg theory change in the expected manner as the 
electron band-filling changes. Note that any sharp features in the normal electronic 
density of states have their effects reduced by the `averaging' over a large phonon 
frequency range. Hence the strong fall in $T_{c}$ and $\Delta$ is expected at both 
small and large fillings ($n<0.45$ and $n>1.55$) due to the rapid fall in the density 
of states in our model. More subtle, is the result that the chemical potential shifts 
by a considerable amount between the normal and superconducting states, if it lies 
near a van Hove singularity. To observe such an effect, the superconductor would have 
to be coupled to one with a more constant density of states.

We find, in agreement with 
previous work~\cite{Free1,Free2,Nico,Free4}, 
that vertex corrections lead to results that correspond to a reduced 
effective strength of the electron-phonon coupling near half-filling, but an increased 
coupling strength near the band edges. These effects are exemplified by reductions in 
critical temperature, $T_{c}$, superconducting gap, $\Delta$, 
isotope coefficient, $\alpha$, and thermodynamic critical field, $H_{c}$, near 
half-filling. As nearly all of these effects can be modeled by an appropriate 
Coulomb pseudopotential, $\mu^{*}_{C}$, it makes it extremely difficult for any single 
experiment to reveal that vertex corrections have played a significant role. However, 
we do find some trends worth pointing out.

Firstly, if there is difficulty in fitting both $T_{c}$ and $\Delta$ with a given 
$\alpha^{2}F(\omega)$ and $\mu^{*}_{C}$, then this is an indication that vertex 
corrections may contribute, since they affect the ratio $2\Delta/kT_{c}$ for fixed 
$T_{c}$, reducing it near half-filling. Secondly, if the experimentally measured 
deviation function for the thermodynamic critical field lies below the predicted 
value [with a given $\alpha^{2}F(\omega)$ and $\mu^{*}_{C}$] vertex corrections could 
be important. Thirdly, a theoretical prediction, ignoring vertex corrections, will 
overestimate the specific heat jump at $T_{c}$. 
Finally, a more striking result, is in the isotope coefficient, 
$\alpha$, which vertex corrections reduce much more markedly than does $\mu^{*}_{C}$. 
Indeed, vertex 
corrections can lead to $\alpha <0$, which $\mu^{*}_{C}$ alone can never 
do~\cite{Mars}. 
Hence materials that have moderate to large $T_{c}$s, but small isotope coefficients 
can still be electron-phonon mediated superconductors with vertex corrections 
included. As the isotope coefficient is the single experimental quantity affected the 
most by vertex corrections, it is important to consider materials which have $\alpha$ 
unexplained by Migdal-Eliashberg theory.

Anomalously low, and even negative isotope coefficients have been 
observed in materials, such as Ru~\cite{Park}, 
$\alpha$-uranium~\cite{Ur}, PdH~\cite{PdH}, 
and La$_{2-x}$Sr$_{x}$CuO$_{4}$~\cite{LSCO,LSCO1,LSCO2}, 
where vertex corrections are probably not important, and other mechanisms, such as 
anharmonicity, conduction electron density variations and paramagnetic impurities 
play a role. However, a system such as Rb$_{3}$C$_{60}$~\cite{C60a,C60b}, 
where the phonon frequency is a 
sizable fraction of the electron bandwidth, is much more likely to have vertex 
corrections affect the value of $\alpha$.

Still, the best way to see the effects of vertex corrections is to directly view 
their contribution in the multiphonon region of a tunnel junction, 
or in the high-energy region of the optical conductivity. 
So, in order to unequivocally demonstrate the presence of strong vertex corrections in a 
material, more accurate dynamical measurements at energies beyond the 
highest phonon frequencies need to be carried out. 

\section{Acknowledgements}

JKF and PM would like to thank the Office of Naval Research for support, under 
the Young Investigator Program Award (N000149610828). 
EJN is a Cottrell Scholar of Research Corporation
and is also supported by the Natural Sciences and Engineering Research
Council of Canada.
We are grateful for useful discussions with J. W. Serene.

\appendix

\section{Conserving Approximation Formula}

\begin{figure}
  \centerline{\psfig{figure=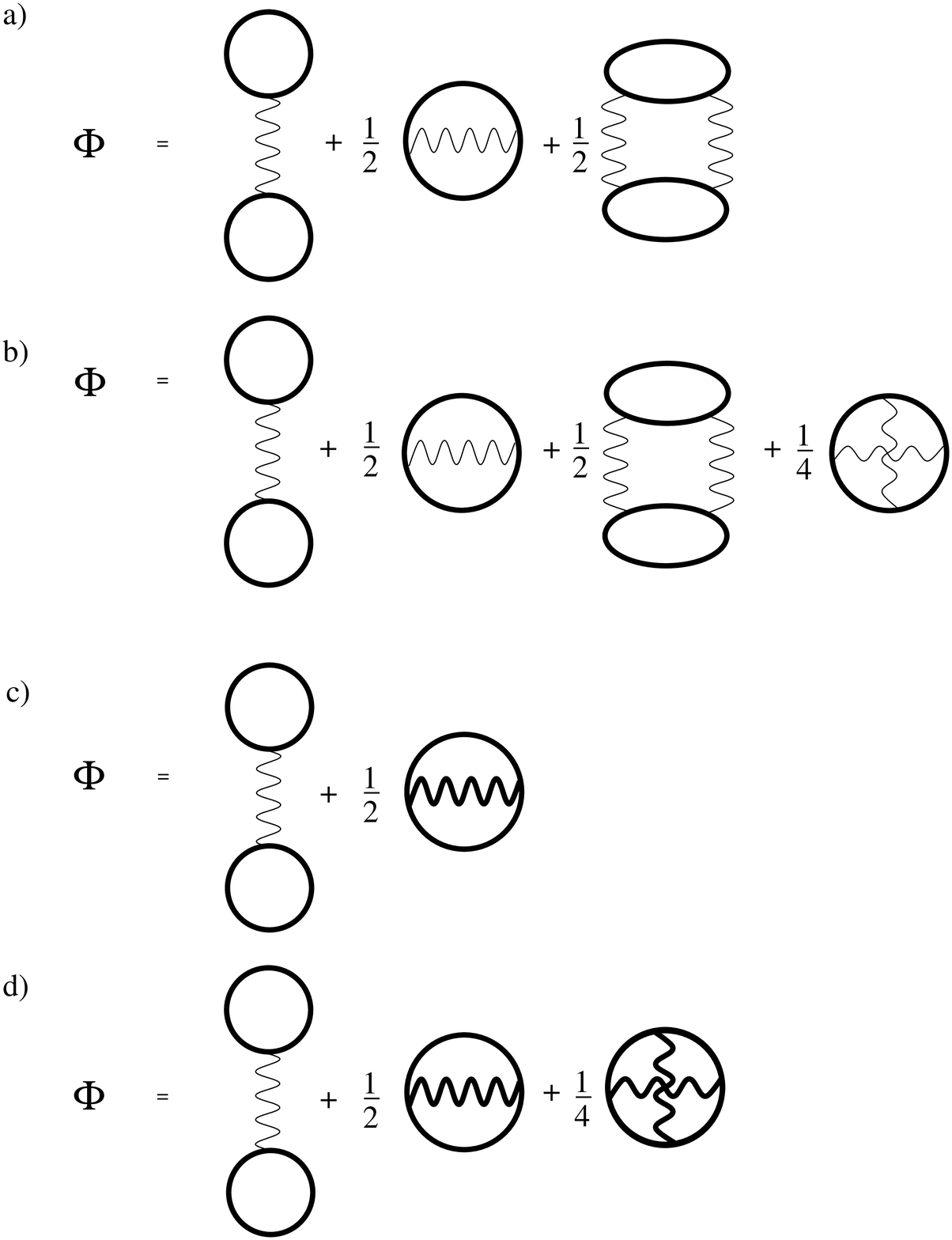,width=4.5in}}
\caption{Free-energy functional, $\Phi$ for the conserving approximations. 
Equations (a) and (b) are with bare phonon propagators, while (c) and (d) have 
dressed phonons. The first term on the right in each equation is the Hartree term, 
which does not change and which is not included in $\Phi'$. 
(Its phonon line is always undressed, to avoid double counting). 
The extra, final diagram in (b) and (d) is the vertex-correction term.}
\label{fig:free}
\end{figure}

\begin{figure}
  \centerline{\psfig{figure=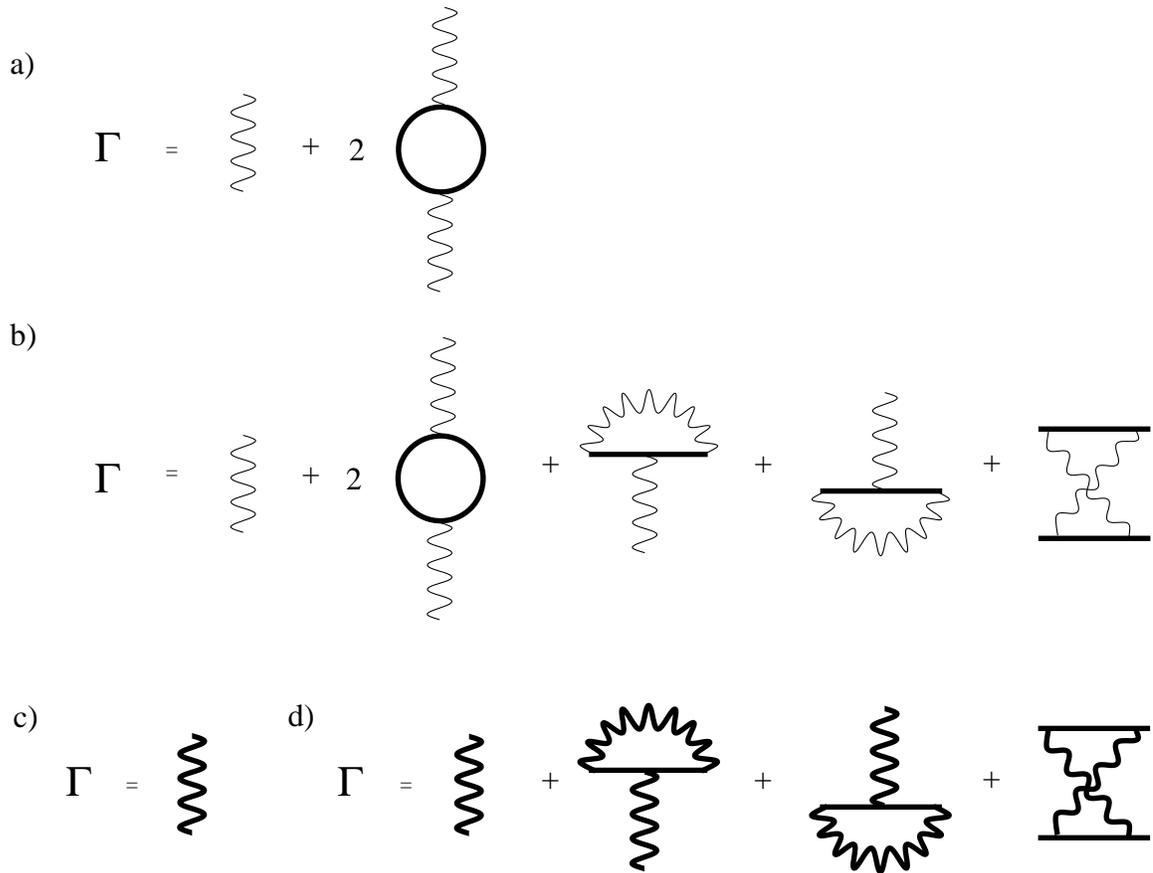,width=6.0in}}
\caption{The irreducible vertex-function diagrams, for superconductivity. 
(a) is with no vertex corrections, 
while (b) includes the three 
vertex-corrected diagrams. When the phonons are dressed, 
(c) is without vertex corrections, while (d) includes them.}
\label{fig:vert}
\end{figure}

In this appendix, we give the specific formulae for the 
free-energy functional, self-energy 
and irreducible vertex functions for each of the four conserving approximations. 
Figures~\ref{fig:free}-\ref{fig:vert} are 
the representations of these equations as Feynman diagrams. 
Hereafter, we employ the shortened notation $G_{n} \equiv G(i\omega_{n})$, 
$G^{*}_{n} \equiv G(-i\omega_{n})$ and similarly for $F_{n}$, $D_{\nu}$ and 
$\pi^{(0)}_{\nu}$. 
Note that the difference of two fermionic frequencies leads to a 
bosonic frequency, {\it i.e.} 
$D_{m-n}= D(i\omega_{m}-i\omega_{n}) = D(i\omega_{\nu})$ 
where $\nu = m-n$. 

The calculations with no vertex corrections and a `bare' phonon propagator have the 
free-energy functional, 
\begin{eqnarray}
\Phi^{bare} & = & \frac{-UT^{2}}{2} \sum_{n,m} 
{\rm Tr} [ \underline{\underline{\tau}}_{3} \, 
\underline{\underline{G}}_{n} ] \,
{\rm Tr} [ \underline{\underline{\tau}}_{3} \, 
\underline{\underline{G}}_{m} ] 
D^{(0)}(\omega = 0) \\* \nonumber
& & 
+ \frac{UT^{2}}{2} \sum_{n,m} 
{\rm Tr} [  \underline{\underline{\tau}}_{3} \, 
\underline{\underline{G}}_{n} \,  
\underline{\underline{\tau}}_{3} \, \underline{\underline{G}}_{m}] 
D^{(0)}_{n-m}
+  \frac{U^{2}T}{4}  \sum_{\nu} \left[ \pi^{(0)}_{\nu} 
 D^{(0)}_{\nu} \right]^{2},
\label{eq:freea}
\end{eqnarray}
where $\pi^{(0)}_{\nu}$ 
is the electron polarizability, defined in Eq.~(\ref{eq:chi}). 
The inclusion of the $\underline{\underline{\tau}}_{3}$ matrices,
ensures that each 
pair of off-diagonal Green's functions, 
$F^{*}_{n}F_{m}$, 
corresponding to Cooper pair creation then annihilation, 
enters the product with a minus sign. 

Functional differentiation with respect to the diagonal electron Green's 
function, $G_{n}$, and off-diagonal Green's function, 
$F^{*}_{n}$ leads 
respectively to the diagonal term in the self-energy, $\Sigma (i\omega_{n})$:
\begin{equation}
\Sigma^{bare} (i\omega_{n}) = UTn + UT \sum_{m} 
G_{m} D^{(0)}_{n-m}
- U^{2}T \sum_{m}  G_{m} \pi^{(0)}_{n-m}
 \left[ D^{(0)}_{n-m} \right]^{2},
\label{eq:selfa}
\end{equation}
and the off-diagonal term, $\phi (i\omega_{n})$:
\begin{equation}
\phi^{bare} (i\omega_{n}) =  -UT \sum_{m} F_{m} D^{(0)}_{n-m}
- U^{2}T \sum_{m}  F_{m} \pi^{(0)}_{n-m}
 \left[ D^{(0)}_{n-m} \right]^{2}.
\label{eq:phia}
\end{equation}

We only require the superconducting vertex part, which is given by the derivative: 
\begin{equation}
T\Gamma^{bare}_{n,m} = \delta \phi(i\omega_{n}) /  \delta F_{m},
\end{equation} 
taken in the limit $F_{m} \mapsto 0$. 
Hence with bare phonons and no vertex corrections, the vertex function is that 
shown in Figure~\ref{fig:vert}(a):
\begin{equation}
\Gamma^{bare}_{n,m} = -U D^{(0)}_{n-m} 
- U^{2} \pi^{(0)}_{n-m}  \left[ D^{(0)}_{n-m} \right]^{2}.
\label{eq:verta}
\end{equation}

There is only one extra term which comes from the inclusion of 
vertex corrections in the free-energy 
functional. It is the diagram in Figure~\ref{fig:free}(b) with crossed phonon lines, 
and is equal to:
\begin{equation}
\Phi^{vc} = \frac{U^{2}T^{3}}{4} \sum_{n,m,l}
{\rm Tr} \left[ 
\underline{\underline{\tau}}_{3} \, 
\underline{\underline{G}}_{n} \, \underline{\underline{\tau}}_{3} \,  
\underline{\underline{G}}_{m} \,
\underline{\underline{\tau}}_{3} \, 
\underline{\underline{G}}_{l} \, \underline{\underline{\tau}}_{3} \,  
\underline{\underline{G}}_{n-m+l} \right]
D^{(0)}_{n-m} D^{(0)}_{l-m}.
\label{eq:freeb}
\end{equation}
The total free-energy functional for bare phonons with vertex corrections included 
is $\Phi^{bare} + \Phi^{vc}$.

The extra term in the self-energy is found by differentiation of the above term. 
For the diagonal and off-diagonal parts, respectively, this leads to:
\begin{eqnarray}
\Sigma^{vc} (i\omega_{n}) & = & U^{2}T^{2} \sum_{m,l} \left\{
G_{m} \left[ G_{l} G_{n-m+l}
-  F_{l} F^{*}_{n-m+l} 
-  F_{n-m+l} F^{*}_{l} \right] \right.
\nonumber
\\*  & & \left.
- G^{*}_{l} F_{m} F^{*}_{n-m+l} \right\}
D^{(0)}_{n-m} D^{(0)}_{l-m},
\label{eq:selfb}
\end{eqnarray}

\begin{eqnarray}
\phi^{vc} (i\omega_{n}) & = & U^{2}T^{2} \sum_{m,l} \left\{
F_{m} \left[ F^{*}_{l} F_{n-m+l}
-  G_{l} G_{n-m+l} 
-  G^{*}_{n-m+l} G^{*}_{l} \right] \right. \nonumber
\\*  & & \left.
-   F_{l} G_{m} G^{*}_{n-m+l} \right\}
D^{(0)}_{n-m} D^{(0)}_{l-m},
\label{eq:phib}
\end{eqnarray}
The extra term leads to the new self-energy, 
$\Sigma = \Sigma^{bare} + \Sigma^{vc}$ and 
$\phi = \phi^{bare} + \phi^{vc}$.

When the vertex correction term is added to the self-energy, 
three new terms appear in 
the irreducible vertex function, coming from each of three Green's functions that 
can be differentiated. The extra diagrams in Figure~\ref{fig:vert}(b) 
contribute a total of:
\begin{equation}
\Gamma^{vc}_{m,n}  =  
U^{2}T \sum_{l} \left\{ \left[ -G_{l} G_{n-m+l} 
-  G^{*}_{n-m+l} G^{*}_{l} \right]
D^{(0)}_{n-m} D^{(0)}_{l-m} 
-  G_{l} G^{*}_{n+m-l} 
D^{(0)}_{n-l} D^{(0)}_{m-l} \right\},
\label{eq:vertb}
\end{equation}
leading to $\Gamma_{m,n} = \Gamma^{bare}_{m,n} + \Gamma^{vc}_{m,n}$ 
as the irreducible vertex function for `bare' phonons, with vertex corrections.

The calculations with no vertex corrections, but a dressed phonon propagator have the 
free-energy functional, 
\begin{eqnarray}
\Phi^{dressed} & = & \frac{-UT^{2}}{2} \sum_{n,m} 
{\rm Tr} [ \underline{\underline{\tau}}_{3} \, 
\underline{\underline{G}}_{n} ] 
{\rm Tr} [ \underline{\underline{\tau}}_{3} \, 
\underline{\underline{G}}_{m}] 
D^{(0)}(\omega = 0) \\* \nonumber
& & 
+ \frac{UT^{2}}{2} \sum_{n,m} {\rm Tr} 
[\underline{\underline{G}}_{n} \underline{\underline{G}}_{m}] 
D_{n-m}.
\label{eq:freec}
\end{eqnarray}

Functional differentiation with respect to the diagonal electron Green's function, 
$G_{n}$, and off-diagonal Green's function, $F^{*}_{n}$ 
leads respectively to the diagonal term in the electron self-energy, 
$\Sigma^{dressed} (i\omega_{n})$:
\begin{equation}
\Sigma^{dressed} (i\omega_{n}) = Un + 
UT \sum_{m} G_{m} D_{n-m}
\label{eq:selfc}
\end{equation}
and the off-diagonal term, $\phi^{dressed} (i\omega_{n})$:
\begin{equation}
\phi^{dressed} (i\omega_{n}) =  -UT \sum_{m} F_{m} D_{n-m}.
\label{eq:phic}
\end{equation}
Similarly, differentiation with respect to the dressed phonon propagator leads to the 
phonon self-energy, 
\begin{equation}
\Pi^{(1)}(i\omega_{\nu}) = -2UT \sum_{m} \left[ G_{m} G_{m+\nu}
-  F_{m} F^{*}_{m+\nu} \right] 
\label{eq:dselfc}
\end{equation}
where the factor of 2 indicates a sum over electron spins.

The superconducting vertex function still retains its simple form, differentiation 
of Eq.~(\ref{eq:phic}) giving:
\begin{equation}
\Gamma^{dressed}_{n,m} = - U D_{n-m} .
\label{eq:vertc}
\end{equation}

Analogously to the case of `bare' phonons, there is only one extra term which 
comes from vertex corrections in the 
free-energy functional. It is equal to:
\begin{equation}
\Phi^{vc2} = \frac{U^{2}T^{3}}{4} \sum_{n,m,l}
{\rm Tr} \left[ \underline{\underline{\tau}}_{3} \, 
\underline{\underline{G}}_{n}  \, 
\underline{\underline{\tau}}_{3} \, \underline{\underline{G}}_{m} \, 
\underline{\underline{\tau}}_{3} \, \underline{\underline{G}}_{l}  \, 
\underline{\underline{\tau}}_{3} \, \underline{\underline{G}}_{n-m+l} 
\right]
D_{n-m} D_{l-m},
\label{eq:freed}
\end{equation}
where now the total free-energy functional for dressed phonons with vertex 
corrections included is $\Phi^{dressed} + \Phi^{vc2}$. 

\begin{figure}
  \centerline{\psfig{figure=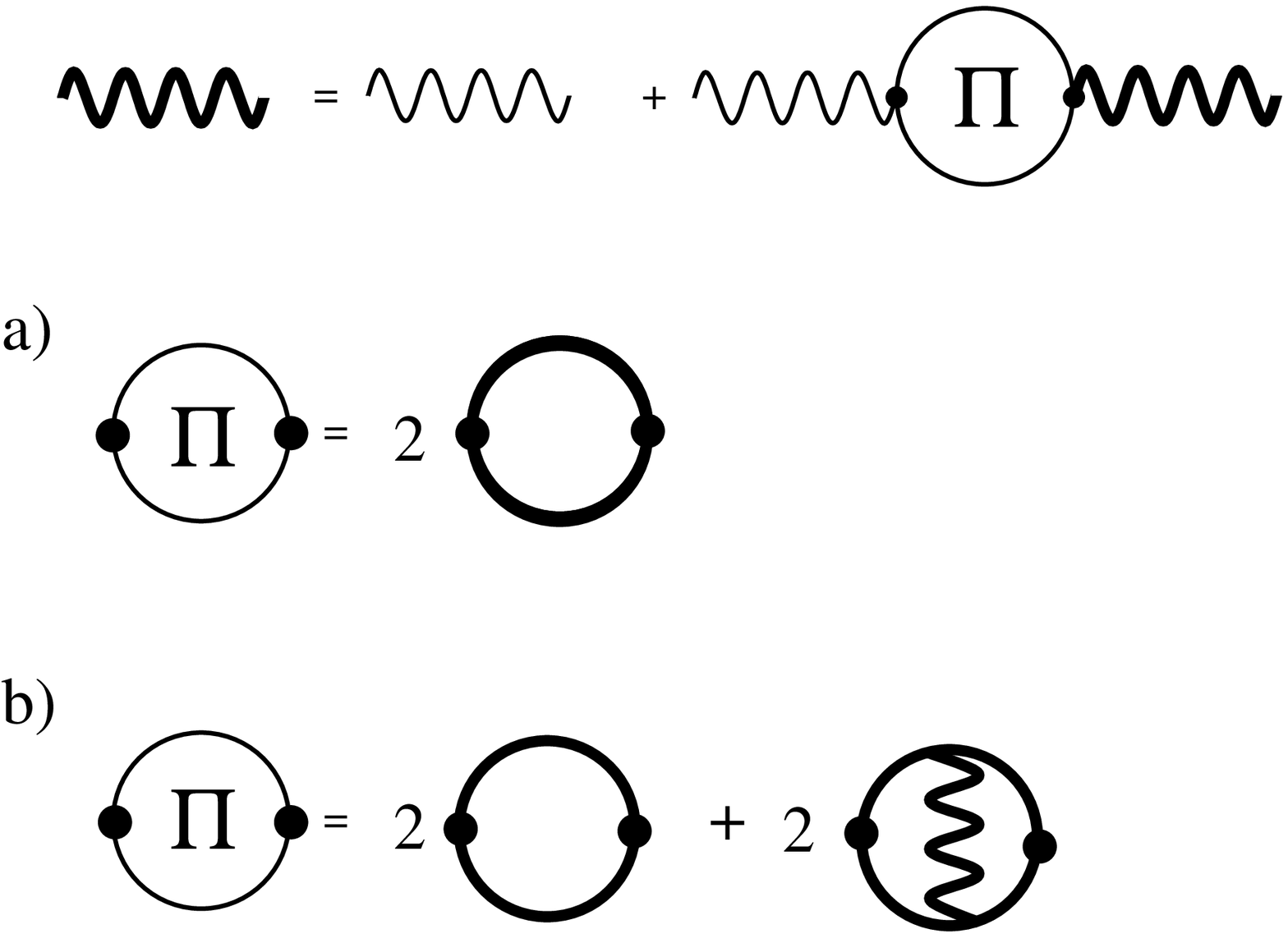,width=4.5in}}
  \caption{Dyson's equation for the phonon propagators.
   Thin lines indicate bare Green's functions while thick lines indicate 
   dressed ones. (a) is the phonon self-energy without vertex corrections, 
   (b) is with vertex corrections. The factors of 2 come from a sum over electron 
   spins. }
\label{fig:dself}
\end{figure}

Differentiation of the above contribution now leads to an extra term in the phonon 
self-energy, $\Pi^{vc}$, as well as the extra electronic self-energy terms. 
$\Pi^{vc}$ is the term shown with crossed phonon lines 
in Figure~\ref{fig:dself}(b):
\begin{equation}
\Pi^{vc}(i\omega_{\nu}) = 
-U^{2}T^{2} \sum_{m,l} Tr \left\{ 
\underline{\underline{\tau}}_{3}\,
\underline{\underline{G}}_{m+\nu}\,
\underline{\underline{\tau}}_{3}\,
\underline{\underline{G}}_{m}\,
\underline{\underline{\tau}}_{3}\,
\underline{\underline{G}}_{l}\,
\underline{\underline{\tau}}_{3}\,
\underline{\underline{G}}_{l+\nu}
\right\}
D_{l-m}.
\label{eq:dselfd}
\end{equation}
Hence the full phonon self-energy is now 
$\Pi(i\omega_{\nu}) = \Pi^{(1)}(i\omega_{\nu}) + \Pi^{vc}(i\omega_{\nu})$. 

The vertex functions and self-energies for electrons gain terms 
equivalent to those in Eqs.~(\ref{eq:selfb}-\ref{eq:vertb}) only with 
bare phonon propagators, $D^{(0)}_{\nu}$, replaced by dressed ones, $D_{\nu}$. 

In the normal state, the expressions for the free-energy functionals and 
self-energies are simplified, by setting all 
off-diagonal contributions to zero, $F_{n}=0$ and $\phi(i\omega_{n})=0$. 
Note that the above formulae for irreducible vertex functions are only calculated 
in the normal state.

%

%

\end{document}